\documentclass[12pt,tightenlines,showpacs,preprintnumbers,
superscriptaddress,nofootinbib]{revtex4}

\usepackage[pdftex]{graphicx}
\newcommand{\PRE}[1]{{#1}} 
\usepackage{bm}
\usepackage{subfigure}
\usepackage{slashed}
\usepackage{epsfig}
\newcommand{\M}{{\cal M}}

\newcommand{\GeV}{\rm GeV}
\def\beq{\begin{eqnarray}}
\def\eeq{\end{eqnarray}}
\def\bea{\begin{eqnarray}}
\def\eea{\end{eqnarray}}
\def\be{\begin{equation}}
\def\ee{\end{equation}}

\newcommand{\gev}{\text{GeV}}

\newcommand{\cm}{\text{cm}}

\newcommand{\s}{\text{s}}

\newcommand{\eqref}[1]{Eq.~(\ref{#1})}

\newcommand{\gsim}{\lower.7ex\hbox{$\;\stackrel{\textstyle>}{\sim}\;$}}
\newcommand{\lsim}{\lower.7ex\hbox{$\;\stackrel{\textstyle<}{\sim}\;$}}

\newcommand{\LH}{\Psi}
\newcommand{\SH}{\Phi}
\newcommand{\mdm}{M_{\rm dm}}

\newcommand{\sv}{\sigma v}
\newcommand{\sveff}{\langle\sigma v\rangle}

\newcommand{\sgll}{\sigma_{\gamma l l}}
\newcommand{\svgg}{\langle\sigma v\rangle_{\gamma\gamma}}
\newcommand{\svgZ}{\langle\sigma v\rangle_{\gamma Z}}
\newcommand{\svgll}{\langle\sigma v\rangle_{\gamma ll}}
\newcommand{\svll}{\langle\sigma v\rangle_{ ll}}
\begin{document}
\preprint{ULB-TH/13-09}



\title{ \PRE{\vspace*{1.5in}}Scalar Dark Matter  Models \\
with Significant Internal Bremsstrahlung
\PRE{\vspace*{0.3in}} }


\author{Federica Giacchino\PRE{\vspace*{.2in}}}
\affiliation{Service de Physique Th\'eorique\\
 Universit\'e Libre de Bruxelles\\ 
Boulevard du Triomphe, CP225, 1050 Brussels, Belgium\PRE{\vspace*{.2in}}}

\author{Laura Lopez-Honorez}
\affiliation{Theoretische Natuurkunde\\
Vrije Universiteit Brussel and The International Solvay Institutes\\
Pleinlaan 2, B-1050 Brussels, Belgium\PRE{\vspace*{.2in}}}

\author{Michel H.G. Tytgat}
\email{federica.giacchino@ulb.ac.be, llopezho@vub.ac.be, mtytgat@ulb.ac.be}
\affiliation{Service de Physique Th\'eorique\\
 Universit\'e Libre de Bruxelles\\ 
Boulevard du Triomphe, CP225, 1050 Brussels, Belgium\PRE{\vspace*{.2in}}}


\begin{abstract}
There has been interest recently on particle physics models that may
give rise to sharp gamma ray spectral features from dark matter
annihilation. Because dark matter is supposed to be electrically
neutral, it is challenging to build weakly interacting massive
particle models that may accommodate both a large cross section into
gamma rays at, say, the Galactic center, and the right dark matter
abundance.  In this work, we consider the gamma ray signatures of a
class of scalar dark matter models that interact with Standard Model
dominantly through heavy vector-like fermions (the vector-like
portal). We focus on a real scalar singlet $S$ annihilating into
lepton-antilepton pairs. Because this two-body final-state
annihilation channel is d-wave suppressed in the chiral limit,
$\sigma_{f\bar f} v \propto v^4$, we show that virtual internal
bremsstrahlung emission of a gamma ray gives a large correction, both
today and at the time of freeze-out.  For the sake of comparison, we
confront this scenario to the familiar case of a Majorana singlet
annihilating into light lepton-antilepton pairs, and show that the
virtual internal bremsstrahlung signal may be enhanced by a factor of
(up to) two orders of magnitude. We discuss the scope and possible
generalizations of the model.
\end{abstract}

\pacs{95.35.+d, 12.60.Jv}
 \maketitle

\newpage

\section{Introduction}

The nature of Dark Matter (DM), which is supposed to account for about 80 \% of all mass in the universe, is one of the big mysteries of physics. It is also  one of the strongest indication for possible physics beyond the Standard Model (SM) of particle physics. Indeed the dominant view is that dark matter is made of new, neutral and stable (or very long-lived) particles. Among the plethora of possible DM candidates, weakly interacting massive particles or WIMPs have many attractive features. First and foremost their abundance may be naturally explained through thermal freeze-out, a mechanism that is very robust, insensitive to unknown, higher scale physics, and points to an almost unique prediction for the annihilation cross section of DM in the early universe, $\langle \sigma v\rangle \sim 10^{-26} \cm^3\cdot \s^{-1}$. This feature also paves the way for the strategies for the identification of DM, provided that DM annihilates into --and thus interact with-- SM degrees of freedom: at colliders, through direct detection, using low background detectors, or through indirect detection, which is the topic of the present work. 

Indirect detection rests on the possibility that DM, which supposedly is accumulated in various parts of the universe, to begin with the central region of our own galaxy, may annihilate into SM particles or messengers, thus contributing to the cosmic flux of particles that reach the Earth or its vicinity. One important issue with this search strategy is that the bulk of the cosmic rays is expected to be of astrophysical origin and thus is somewhat uncertain. This, combined with the fact that the spectral energy distribution of messengers produced from DM annihilation is generically featureless, somewhat limits our ability to non-ambiguously identify DM -- of course we may use the data to set exclusion limits and, in practice we do so, since there is no yet any clear signal of DM from the sky. 

Possible exceptions to this rule of thumb is offered by so-called smoking guns, that is  signatures that in principle have no counterpart of astrophysical origin. An important instance for our purpose is a gamma ray line (i.e. a monochromatic photon) from DM annihilation into $\gamma\gamma$ \cite{Bergstrom:1988fp,Rudaz:1989ij} or $Z \gamma$ \cite{Urban:1992ej} (see~\cite{Bringmann:2012ez} for a recent review). Gamma ray lines are actively being searched, most notably by the Fermi satellite~\cite{Fermi-LAT:2013uma} and the HESS telescope~\cite{Abramowski:2013ax}, and again the absence of signal has so far only permit to set exclusion limits. Since DM is neutral, its annihilation in $\gamma\gamma$ should proceed through radiative corrections, and so is  {\em a priori} suppressed by a factor $\propto \alpha^2 \lesssim 10^{-4}$ compared to  annihilation into fermion or gauge boson pairs, $f\bar f$ or $WW$,  which not only are supposed to determine the relic abundance but also lead to a large $\gamma$ ray continuum. However this is not a no-go theorem and following the recent claim of a possible excess of gamma rays around $E_\gamma \sim 130 \;\gev$ in the  Fermi telescope data~\cite{Bringmann:2012vr,Weniger:2012tx}, much works have been dedicated to find new ways to circumvent  this conclusion \cite{Dudas:2012pb,Cline:2012nw,Ibarra:2012dw,Kyae:2012vi,Buckley:2012ws,Chu:2012qy,Das:2012ys, Weiner:2012cb,Tulin:2012uq,Acharya:2012dz} (see also~\cite{Gustafsson:2007pc,Mambrini:2009ad,Jackson:2009kg,Arina:2009uq, SchmidtHoberg:2012ip,Wang:2012ts}). Despite the apparent fading of the significance of the signal~\cite{Fermi-LAT:2013uma}, we believe that it remains of interest to look for further alternative scenarios. 

In the present work, we specifically focus on virtual internal
bremsstrahlung (VIB) as a possible way of producing an enhanced, sharp
gamma ray spectral feature from DM
annihilation~\cite{Bergstrom:1989jr,Flores:1989ru}. VIB is a process
by which a gamma ray in the final state is, roughly speaking, emitted
by a charged virtual particle \footnote{Distinction from of soft
  photons through final state radiation (FSR) may be made in a gauge
  invariant way~\cite{Bringmann:2007nk}.}. While being suppressed by a
factor of $\alpha$, it may actually dominate DM annihilation if the
two-body process is suppressed. The canonical example is the case of
two Majorana particles $\chi$ annihilating into a pair of light
fermions $f\bar f$~\cite{Bergstrom:1989jr,Flores:1989ru}. Because of
Pauli principle, annihilation of the Majorana particles must be either
in a s-wave spin 0 state or p-wave spin 1. In the chiral limit
$m_f=0$, the latter case is the only possibility as the effective
coupling between the pair of $\chi$ and the $f\bar f$ is of the
pseudo-vector type (which implies that the final state is $J=1$). So
annihilation of the Majorana pair is either chirally or p-wave
suppressed, $\sigma_{f\bar f} v \propto m_f^2$ or $\propto v^2$
respectively~\cite{Goldberg:1983nd}. While of little practical
importance for freeze-out in the early universe, for which $v^2 \sim
0.24$, the suppression is dramatic at the galactic center, where $v
\sim 10^{-3}$. On the contrary, the emission of a photon in the final
state allows for $J = 0$, so 3-body annihilation may proceed in the
s-wave channel, which is by far the dominant process at the galactic
center.

This beautiful mechanism has been first proposed within the framework of the MSSM~\cite{Bergstrom:1989jr}, but since then has been considered and studied further in more general terms (see e.g.~\cite{Baltz:2002we,Bringmann:2007nk,Ciafaloni:2011sa,Bell:2011if,Barger:2011jg,Garny:2011ii}). In particular,  a very simple scenario has been proposed in~\cite{Bergstrom:2012bd} in an attempt to explain the putative 130 GeV Fermi excess. The dark matter candidate is a right-handed neutrino, and the signal is annihilation into lepton-antilepton pairs together with a gamma. In this work, it has been shown that both the relic abundance and the Fermi  measurement could be simultaneously explained, assuming a slight ${\cal O}(10)$ astrophysical boost of the gamma ray signal. In the present work, we consider a scalar dark matter candidate instead, with properties which are otherwise very similar to those of the heavy neutrino of~\cite{Bergstrom:2012bd}, hence we will adopt this instance as a benchmark to which to compare our model. The basic facts we will use is that the annihilation of a real scalar DM candidate into fermion-antifermions pair is either s-wave but chirally suppressed $\sigma_{f\bar f} v \propto m_f^2$, like the Majorana case, or d-wave suppressed, $\sigma_{f\bar f} v \propto v^4$. The extra velocity suppression compared to the Majorana case may seem harmless if we consider annihilation in the early universe, but will show otherwise. In particular we will show that VIB annihilation may give to very significant annihilation into gamma rays for candidates that match the measured cosmic abundance. For the sake of the argument we will consider a very simple toy model, and limit ourselves to a leptophilic scalar DM candidate.

We begin with a discussion of the basic features of the model, including its annihilation into lepton-antilepton pairs. Next we give some details of our calculation of its 3-body annihilation in the VIB channel and compare the results with the Majorana case. In particular we show that the VIB signal is strongly enhanced in the scalar case compared to the Majorana case. 
In the last section we discuss the possible generality of this results, possible drawbacks, and prospects. We finish with some conclusions.


\section{The model}
\label{sec:model}

The  model we consider is very simple. It consists of a real scalar particle, $S$, which we take to be leptophilic for the sake of our argument. By this we mean that it has Yukawa couplings only to SM leptons. More specifically, in this section we consider couplings to the right-handed ones ($l_R$). We will discuss other possibilities in Sec.~\ref{sec:prospects}. We also introduce  heavy vector-like leptons ($\LH$). At this stage it does not matter whether there is one or many (like one per SM generation) heavy leptons. Their Yukawa interactions are thus of the form 
\begin{equation}
  \label{eq:yuk1}
{\cal L} \supset y_l\; S\; \bar \LH l_R + h.c.\,.
\end{equation}
In this specific instance the ${\LH}$ are thus $SU(2)_L$ singlet and obviously their hypercharges are equal $Y_{\LH} = Y_{l_R}$. We want to consider the possibility that the $S$ is a dark matter candidate. To insure its stability, we assume that the full Lagrangian is invariant under a discrete $Z_2$ symmetry, 
$$
S \longrightarrow - S$$
and 
$$ \LH \longrightarrow - \LH$$ while all the SM fields are taken to be
even under $Z_2$. We will further assume that the $Z_2$ symmetry is
unbroken and the lightest odd particle is the scalar, $M_{\LH} >
M_S$. Hence, in this model, the annihilation of DM into SM fields goes
through heavy vector-like fermions exchange.  Such an interaction has already been
considered in~\cite{Boehm:2003ha} when studying scalar dark matter
candidates in the MeV range.  Following~\cite{Perez:2013nra}, we call this scenario the vector-like
portal.  Notice that, unlike in~\cite{Perez:2013nra}, the scalar
field is taken to be a real singlet, like in other simple models of
non-fermionic dark
matter~\cite{Silveira:1985rk,Veltman:1989vw,McDonald:1993ex,Burgess:2000yq}. Being
a scalar singlet $S$ has also renormalizable coupling to the SM scalar
({\em aka} the Higgs) portal,
\begin{equation}
  {\cal L} \supset {\lambda_S\over 2} S^2 \vert H \vert^2\,.
\label{eq:ls}
\end{equation}
We begin by assuming that this coupling is subdominant, but we comment
on relaxing this condition in Sec.~\ref{sec:prospects}.


\subsection{Two-body annihilation }
This being laid down, we now consider the annihilation process
\begin{equation}
  S(k_1)\; S(k_2) \rightarrow l (p_1)\;\bar l(p_2).
\label{eq:Sann}
\end{equation}
For reasons that will become clear, we give a rather pedestrian
derivation of a the annihilation amplitude.
\begin{figure}
\begin{center}
\includegraphics[width=10cm]{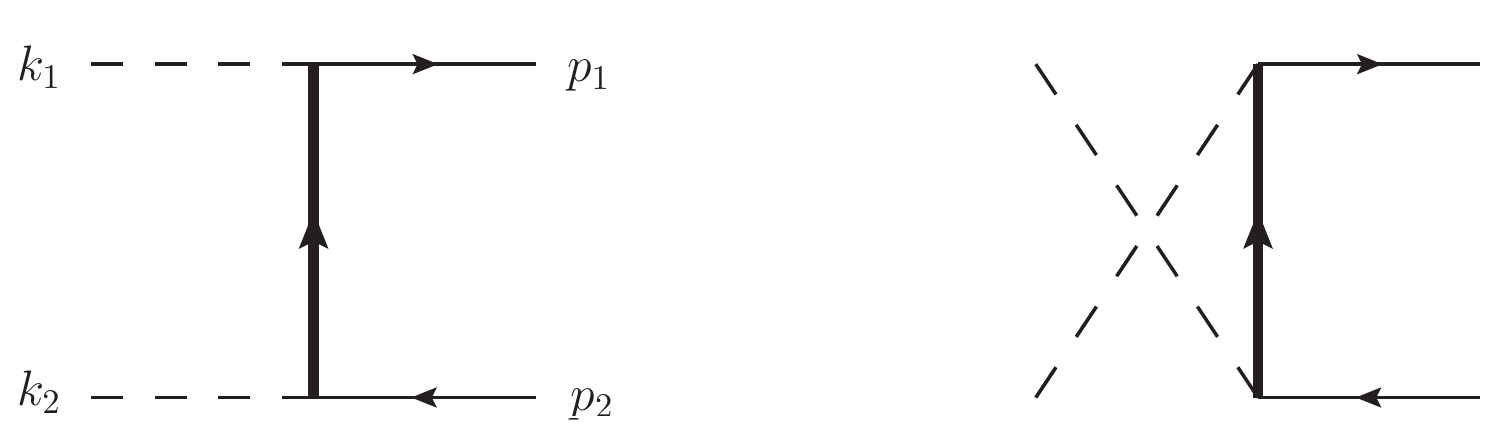}
\end{center}
\caption{Diagrams contributing to the annihilation amplitude (t and u
  channels) through the vector-like portal.}
\label{fig:scalar_ff}
\end{figure}
From Fig.~\ref{fig:scalar_ff} the $t$ and $u$ channels amplitudes are given by
$$
 {\cal M}^{(t)}_{f \bar f} =  y_l^2\, \bar u(p_1)P_L {1 \over  \slashed{p}_1 - \slashed{k}_1 - M_{\LH}} P_R v(p_2)
$$
and
$$
 {\cal M}^{(u)}_{f \bar f} =  y_l^2\, \bar u(p_1)P_L {1 \over  \slashed{p}_1 - \slashed{k}_2 - M_{\LH}} P_R v(p_2)
$$ respectively where $P_{R(L)}$ is the projector on right (left)
 helicity. The total amplitude reads
$$
{\cal M}_{f\bar f} = {1\over 2} y_l^2\; \bar u(p_1) P_L\left[(\slashed{k}_1 - \slashed{k}_2) (D_{11} - D_{22}) + (\slashed{p}_2 - \slashed{p}_1) (D_{11} + D_{21})\right] v(p_2)
$$
where 
$$
D_{ij} = {1\over (k_i - p_j)^2 - M_{\LH}^2}.$$
Here and in the next section we adopt the notations of~\cite{Ciafaloni:2011sa}. 
Using the equations of motion for the fermion and the antifermion, this becomes
\begin{equation}
  \label{eq:amp1}
{\cal M}_{f\bar f} = {1\over 2} y_l^2\; \bar u(p_1)
\left[P_L(\slashed{k}_1 - \slashed{k}_2) (D_{11} - D_{21}) - m_f
  (D_{11} + D_{21})\right] v(p_2)\,.
\end{equation} 
The first term is velocity
suppressed, while the second term is proportional to the fermion
mass. Neglecting terms with powers of  $ m_f$ larger than one,
using
\begin{eqnarray}
D_{11} - D_{21} &=& D_{11} D_{21} \times 2 p_1\cdot(k_2- k_1)
\end{eqnarray}
and 
\begin{eqnarray}
D_{11} + D_{21} &= & D_{11} D_{21} \left(2 M_{S}^2 - 2 M_{\LH}^2 - 2 p_1\cdot p_2\right)\, ,
\end{eqnarray}
we get
\begin{equation}
{\cal M}_{f\bar f} = {1\over 2} y_l^2\; \bar u(p_1) \left[P_L(\slashed{k}_2 - \slashed{k}_1) \; 2 p_1\cdot(k_1- k_2)  - m_f \left(2 M_{S}^2 - 2 M_{\LH}^2 - 2 p_1\cdot p_2\right)\right] v(p_2) \;D_{11} D_{21}\,.
\end{equation}
The interesting point is that in the chiral limit, $m_f\rightarrow 0$, the
amplitude squared becomes proportional to $(k_2 - k_1)^2\propto
v^2$, where $v$ denotes the relative velocity of the annihilating $S$
particles. Assuming that $m_f$ negligible and  working in the non-relativistic
limit relevant for annihilation of WIMPs,
we get at leading order in $v$
\begin{equation}
\sigma v(S S \rightarrow l\bar l) = { y_l^4 \over 60 \pi}{v^4\over M_S^2}{1\over (1+r^2)^4}
\end{equation}
for the annihilation of $S$ into a light lepton-antilepton pair. Here
$r$ refers to the ratio of masses $M_{\LH}/M_S$. 

\bigskip 
The suppression by $ v^4$ in the chiral limit is a bit unusual. For the sake of comparison, the corresponding annihilation of a pair of a gauge singlet Majorana into light leptons through a heavy charged scalar ($\SH$), with coupling
\begin{equation}
  {\cal L} \supset g_l \SH^\dagger \chi l_R + h.c.
\label{eq:majDM}
\end{equation}
 is given by
 \begin{equation}
   \sigma v (\chi \chi \rightarrow l\bar l) = {g_l^4 \over 48 \pi}\, {v^2\over M_\chi^2}\, {1+ r^4\over (1+r^2)^4}
\label{eq:Maj2bdy}
 \end{equation}
which shows the usual p-wave suppression $\propto v^2$ and $r =
M_{\SH}/M_\chi$ (see also~\cite{Garny:2011ii} Eq.A.5).  In the
following, $r$ will always refer to the ratio of the mass of the
next-to-lightest particle odd under the $Z_2$ (NLZP) symmetry divided
by the dark matter mass and it is always larger than one. For
identical Yukawa couplings, DM masses and ratios $r$, we have the
following ratio of the averaged cross sections into lepton-antilepton
pairs\be
\label{eq:ratio}
{\svll|_S \over \svll|_\chi } = {4 \, \langle v^4\rangle\over 5\langle
  v^2\rangle (1+r^4)} \lsim 0.16 \ee where the bound is obtained
assuming $v^2 = 0.24$ and $r=1$.  In other words, for identical
DM masses, assuming thermal freeze-out, it takes a larger Yukawa
coupling for a scalar $S$ than for a Majorana $\chi$ to match the
observed relic abundance of DM. This behavior will be studied in more
details in Sec.~\ref{sec:pheno}.

\subsection{Chiral suppression from an effective operator perspective}

The d-wave, $\propto v^4$, suppression of the annihilation cross
section of real scalars in fermion-antifermion pairs is easy to
explain. The annihilation of a pair of scalar in a s-wave corresponds
to a CP even state. Thus, the fermion-antifermion final state must be
described by a CP even scalar bilinear operator e.g. $\bar\psi_f
\psi_f$. However the Yukawa interaction of (\ref{eq:yuk1}) involves
chiral (here right-handed) SM fermions. Hence, the s-wave annihilation
is chirally suppressed. In other words, the amplitude of 
(\ref{eq:amp1}) derives from the dimension 5 operator: \be
\label{eq:effopS}
{\cal O}_S = m_f \, S^2\, \bar l l\,.
\ee
In principle, we could
have annihilation in a p-wave, at least based on the constraint from
CP. However there is no CP-odd bilinear operator involving two
identical {\em real} scalars (i.e. no current) hence the next
possibility is a dimension 8 operator of the following form
\be
\label{eq:effopT}
{\cal O}_T = \partial_\mu S \partial_\nu S\, \Theta_R^{\mu\nu}
\ee
where $\Theta_R^{\mu\nu}$ is the stress-energy tensor of the Dirac field $l_R$,
$$
\Theta_R^{\mu\nu} = {i\over 2} \bar l_R\left(\gamma^\mu \overrightarrow \partial^\nu - \gamma^\nu \overleftarrow \partial^\mu\right)l_R
$$
 Clearly the (traceless part of) $\Theta_R^{\mu\nu}$ has $J=2$ which implies that the annihilation of a pair of real scalars is d-wave suppressed in the chiral limit. 

\bigskip

The tentative conclusion of this section is that 2-body annihilation of a pair of real scalar into SM fermions is suppressed compared to the case of a Majorana particle. We will show that this will have concrete implications. 

\section{Virtual Internal Bremsstrahlung} 
\label{sec:vib}

In this section we turn to radiative processes, and in particular to
internal bremsstrahlung. This is of interest for two reasons.  First,
as it is well-known, the annihilation cross section in the s-wave is
no longer suppressed, which implies that the 3-body final state
process may be more important than the 2-body process, despite the
suppression of the former by a factor ${\cal O}(\alpha/\pi)$. This is
typically the case for annihilation in light fermions (e.g.  leptons)
and when the relative velocity is non-relativistic, like at the
galactic center ($v \sim 10^{-3}$). Second, the emission from
(essentially) the virtual intermediate particle, depending on the
ratio of its mass and that of DM, may show a sharp spectral feature,
which may even mimic a monochromatic gamma ray line. As mentioned in
the introduction, the literature on this topic is vast. In this
section we specifically follow the detailed and pedagogical approach
of~\cite{Ciafaloni:2011sa} to derive the annihilation cross section of
the scalar $S$ to ${\cal O}(\alpha)$. We warn the reader that 1/ the
result is disappointingly simple -- the spectrum of gamma rays and
leptons/antileptons is the same as in the case of a Majorana
particle-- and 2/ this fact was already discussed in the
literature~\cite{Barger:2011jg} (see also~\cite{Boehm:2006df} in the
case of light dark matter). Our excuse to re-iterate is that we
believe that it is of some interest of being a bit more detailed (and
compare to the calculation of the Majorana case of,
say,~\cite{Ciafaloni:2011sa}) and also that the precise normalization
is of prime importance for indirect searches -- so that it is perhaps
worth providing an independent check.

\subsection{Derivation of 3-body cross-section}

\begin{figure}
\begin{center}
\includegraphics[width=16cm]{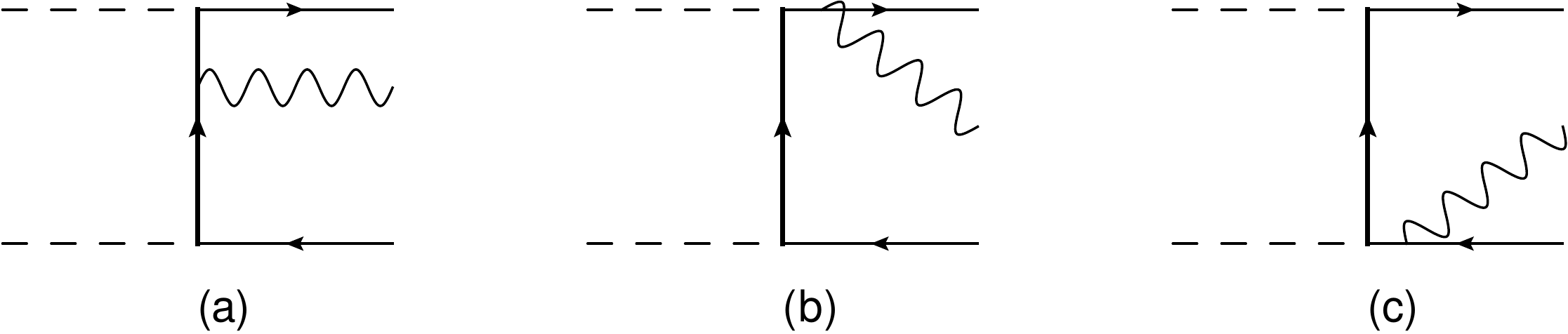}
\end{center}
\caption{Diagrams contributing to the amplitudes for $S S \rightarrow
 \bar l  l \gamma$ process.}
\label{fig:vbi}
\end{figure}

The relevant t-channel diagrams are shown in Fig.~\ref{fig:vbi}. There
are also the three exchange diagrams (u-channel), which add up in the
scalar case (while they subtract in the Majorana case). As before, $
k_1$ and $k_2$ denote the momenta of the annihilating dark matter
particles, $ p_1$ and $p_2$ denote the momenta of the outgoing leptons
and we write $k$ the momentum of the photon,
$$
S(k_1) \, S(k_2) \rightarrow \bar l_R(p_1)\,  l_R (p_2) \,\gamma(k)
$$
and write
$$
{\cal M}\cdot \epsilon^\ast = e y_l^2 \left[ {\cal M}_a^{(t)} + {\cal M}_a^{(u)} + {\cal M}_b^{(t)} + {\cal M}_b^{(u)} + {\cal M}_c^{(t)} + {\cal M}_c^{(u)}\right]\,
$$ where the $t, u$ superscript refers to the
$t,u$-channel diagram. Having in mind s-wave annihilation, we may set $k_1
= k_2 = K$, so that in this limit and after some obvious
manipulations, the first two amplitudes, which are associated to inner
emission, may be written as 
\begin{equation}
{\cal M}_a^{(t)} = \bar u(p_1) P_L \left[(M_{F}^2 + M_S^2) \slashed \epsilon^\ast - 2 \epsilon^\ast \cdot K \slashed K\right] v(p_2) D_{1} D_{2} \equiv {\cal M}_a^{(u)}
\label{eq:Ma}  
\end{equation}
where
$$
D_i = {1\over (p_i - K)^2 - M_{\LH}^2}.
$$

Similarly the final state radiation amplitudes may be written as  
\begin{equation}
  {\cal M}^{(t)}_b = \bar u(p_1) P_L \left[2 \epsilon^\ast\cdot p_1 +  \slashed \epsilon^\ast\slashed k\right] \slashed K v(p_2) D_2 D(p_1 + k) \equiv {\cal M}_b^{(u)}
\label{eq:Mb}
\end{equation}
where the  final state fermion propagator
$$
D(p_1 + k)  = {1\over (p_1 + k)^2 - m_f^2} \equiv {1\over 2 p_1\cdot k}
$$ displays the usual infrared divergent behaviour (here we work in
the chiral limit $m_f=0$), and
\begin{equation}
  {\cal M}_c^{(t)}   = \bar u(p_1) P_L \slashed K\left[ 2 p_2 \cdot \epsilon^\ast + \slashed k \slashed \epsilon^\ast\right] v(p_2) D_1 D(p_2 + k)\equiv {\cal M}_c^{(u)}\,.\label{eq:Mc}
\end{equation}
Now it is easy to see that the potentially IR divergent pieces cancel
from the FSR (Eqs.~(\ref{eq:Mb}) and~(\ref{eq:Mc})) amplitudes. Using
$2 K = p_1 + p_2 + k$ and the equation of motion $\bar u(p_1) \slashed
p_1 = 0$, $\slashed p_2 v(p_2)$ and $k^2 = 0$, the ${\cal M}_b$
amplitude reduces to
\begin{eqnarray}
   {\cal M}_b&=& \bar u(p_1) P_L \left[ 2 \epsilon^\ast\cdot p_1 \slashed k +
  \epsilon^\ast \slashed k \slashed p_1\right] v(p_2) D_2
D(p_1+k)\nonumber\\ 
&=& \bar u(p_1) P_L \left[ 2 p_1 \cdot k
  \slashed\epsilon^\ast \right] v(p_2)D_2 D(p_1+k) = \bar u(p_1)
P_L \slashed\epsilon^\ast v(p_2)D_2. 
\end{eqnarray}
 Similarly the ${\cal M}_c$ amplitude is simply given by
$$
{\cal M}_c = \bar u(p_1) P_L \slashed\epsilon^\ast v(p_2)D_1.
$$
Using
$$
D_1 + D_2 = -(2 M_{\LH}^2 - 2 M_S^2 + 2 K\cdot(p_1 + p_2)) D_1 D_2\, ,
$$
we may combine the six amplitudes to get
\be
\label{eq:3bodyamp}
{\cal M}_{\rm tot} = \bar u(p_1) P_L \left[ k\cdot (p_1 + p_2) \slashed \epsilon^\ast - \epsilon^\ast\cdot (p_1 + p_2) \slashed k\right] v(p_2) D_1 D_2.
\ee
The total amplitude is manifestly gauge-invariant (${\cal M}_{\rm tot} = 0$ for $k \rightarrow \epsilon^\ast$ as it should be), but the derivation makes clear that both the internal and FSR processes are necessary for this to occur. It is also IR divergence free, as expected on general grounds (see~\cite{Bringmann:2007nk}). Indeed, IR divergences in FSR, which are ${\cal O}(\alpha)$ are expected to cancel with similar contributions from the interference of a tree level amplitude and its ${\cal O}(\alpha)$ one-loop radiative corrections (see e.g.~\cite{Peskin:1995ev}).
In the case at hand, just like in the case  of the annihilation of Majorana particles, the tree level amplitude in an s-wave initial state vanishes in the chiral limit. 

\subsection{ Spectrum and cross section dependence in $\bf r$ for VIB }
\label{sec:spectr-cross-sect}

The amplitude (\ref{eq:3bodyamp}) bears little resemblance to the one
of the Majorana case (see e.g. Eq.(4.12) in~\cite{Ciafaloni:2011sa}). Yet it gives rise to precisely the same
gamma ray spectrum.  Defining the 3-body annihilation cross section as
\begin{eqnarray}
  vd\sigma_{2\rightarrow 3}&=&\frac{|\M|^2}{128\pi^3} dxdy
\label{eq:dsv3}
\end{eqnarray}
where $v={\sqrt{{k_1 \cdot k_2}-m_1m_2}}/{E_1E_2}$ refers to the
relative velocity of the $S$ particles and $x, y$ are the reduced
energy parameters $x=2E_\gamma/\sqrt{s}$ and $y=2E_f/\sqrt{s}$, with $s$
the Mandelstam variable corresponding to the center-of-mass energy
squared, we obtain the following amplitude squared for the 3-body
annihilation with emission of a photon
\begin{equation}
\label{eq:amp3bodyS}
 |\M_S|^2=\frac{32 \pi\, \alpha \,y_l^4 }{ M_S^2}\,\frac{ 4  (1 - y) (2 + 2 x^2 + 2 x ( y-2) - 2 y + y^2) }{
 (1 - r^2 - 2 x)^2 (3 + r^2- 2 x - 2 y)^2}  
\end{equation}
with $r = M_{\LH}/M_S$, in agreement with~\cite{Barger:2011jg}. 

As above, it is of interest to compare this expression to the one
obtained in the same limit in the Majorana case (see~\cite{Barger:2011jg} and also Eq.A.1 in~\cite{Garny:2011ii}) 
\begin{equation}
  \label{eq:amp3bodyChi}
{1\over 4} \sum_{\rm spin}|\M_\chi|^2=\frac{4 \, \pi \,\alpha
  \,g_l^4}{M_\chi^2 } \, \frac{ 4 (1 - y) (2 + 2 x^2 + 2 x( y-2) - 2 y
  + y^2) }{(1 -r^2 - 2 x)^2 (3 +r^2 - 2 x - 2 y)^2}
\end{equation}
     where $r=M_{\SH}/M_\chi$. Clearly the dependence on $x$ and $y$
     are precisely the same. It is interesting to notice that, all
     other things being kept constant, the cross section is larger by
     a factor of 8 in the scalar case compared to the Majorana case. A
     factor of $4$ comes clearly from the spin average. The extra
     factor of $2$ is related to the projection of the Majorana pair
     into a spin zero initial state (i.e.  there is a factor of
     $1/\sqrt{2}$ in the amplitude). Let us emphasize that the
     relevant normalization scalar versus Majorana is not important
     for our argumentation, what matters is 2-body versus 3-body.

\begin{figure}[ht]
  \begin{center}
        \includegraphics[width=10cm]{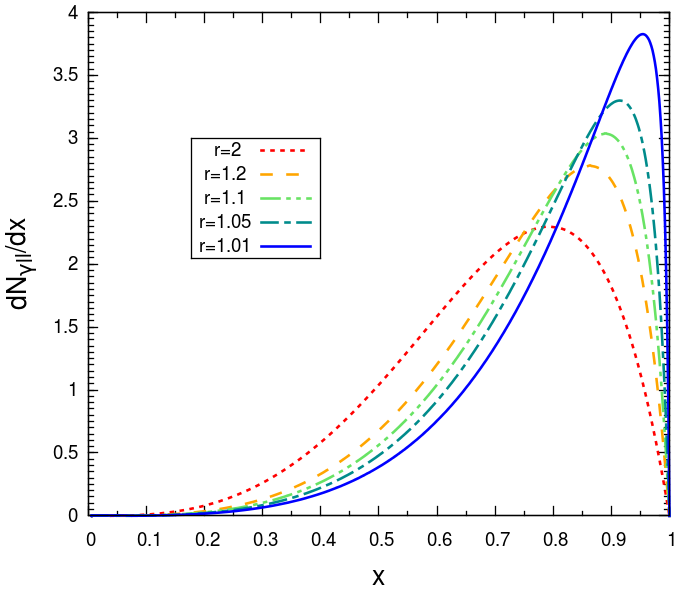}
    \end{center}
    \caption{Spectra $dN_{\gamma ll}/dx=d\log[\sgll]/dx$ as a function of $x=E_\gamma/\mdm$ for several values of $r$. }
  \label{fig:spectr}
\end{figure}

In Fig.~\ref{fig:spectr} we show the dependence of the photon spectrum   
\begin{equation}
  \frac{dN_{\gamma ll}}{dx}=\frac{\mdm}{\sgll}\frac{d \sgll}{dE_\gamma}
\end{equation}
in the parameter $r$.  The spectra have been obtained integrating the
amplitude (\ref{eq:amp3bodyS}) over $\quad (1-x) \leq y = {E_f\over
  \mdm} \leq 1$. We see that for larger values of $r$ the amplitude of
the spectra is lower but the spread is larger. The sharp feature
in the spectrum also moves to smaller $E_\gamma<\mdm$ with increasing
$r$. 

 Let us emphasize that $dN_{\gamma ll}/dx$ is independent of the
scalar/Majorana nature of the dark matter candidate and this
concordance has already been
elucidated in~\cite{Barger:2011jg}.
It is related to the fact that the scalar and Majorana initial states,
while both $L= S= 0$, differ only by their parity, the scalar case
being clearly CP even. We have nothing new to add here, but merely
repeat their argument, which stems from the fact that the amplitude
(\ref{eq:3bodyamp}) may be derived from the effective operator
$$
{\cal O}_S = \left(\partial_\mu \bar l_R \gamma_\nu l_R + \bar l_R \gamma_\nu \partial_\mu l_R\right) F^{\mu\nu}
$$
while in the Majorana case it is related to
$$
{\cal O}_\chi = \left(\partial_\mu \bar l_R \gamma_\nu l_R + \bar l_R \gamma_\nu \partial_\mu l_R\right) \tilde F^{\mu\nu}\, ,
$$ where $\tilde F^{\mu\nu}$ is the dual of $F^{\mu\nu}$, which
only amounts to exchanging the role of the $\vec E$ and the
$\vec B$ of the photon and thus gives rise to the same spectrum.

\bigskip
Although the 3-body cross section $\sgll$ differ for Majorana and
scalar dark matter by numerical factors, we know that the $r$
dependence is universal and, in the appendix, we give a formulation of
$\svgll$ that emphasize this fact. Notice that a useful approximate
expression of $\svgll$ has also been derived previously in~\cite{Bringmann:2007nk}.

\subsection{Annihilation into $\gamma \gamma$ versus  VIB }
\label{sec:annih-into-gamma}

For the sake of completeness, and because of their close relation to internal Bremsstrahlung, we discuss in this section the annihilation of DM in gamma ray pairs. As is well-know, the s-wave annihilation of a pair of Majorana particles in monochromatic gamma rays may be derived from a chiral anomaly argument~\cite{Rudaz:1989ij,Bergstrom:1989jr}. Concretely, in the chiral limit and for $r = {M_\phi/M_\chi} \gg 1$, the amplitude for annihilation may be obtained by simply replacing, in the box diagram corresponding $\chi \chi \rightarrow \gamma \gamma$, the scalar propagator by an effective contact interaction between the Majorana particles and the light fermions. Since this effective coupling is of axial-vector type, the resulting triangular diagram, which only involves light fermions, 
is precisely the one that arises in the derivation of the chiral anomaly. This implies in particular that the amplitude is non-vanishing even for massless fermions.  For $r \gtrsim 1$ and in the chiral limit, the annihilation cross section is simply given by~\cite{Rudaz:1989ij}
$$
\sigma v(\chi \chi \rightarrow \gamma \gamma) = {\alpha^2\over 64 \pi^3} {g_l^4\over M_{\chi}^2}{1\over r^4}.
$$
The general expression, still in the chiral limit but valid for all $r$ (including $r=1$) as obtained from the calculation of the box diagrams~\cite{Bergstrom:1989jr}, is given for reference in  the appendix. 

\begin{figure}
  \includegraphics[width=8cm]{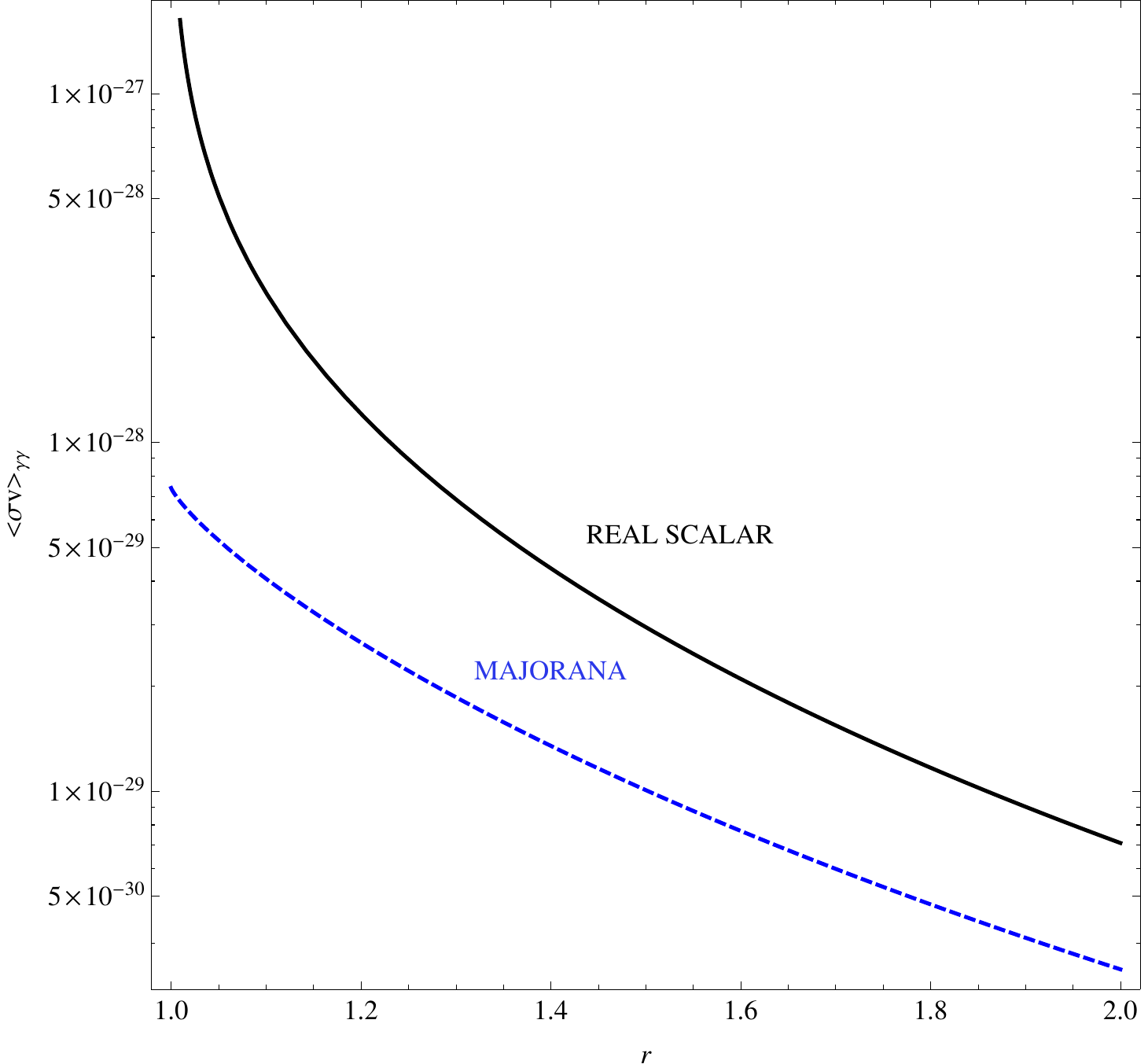}
  \caption{$\svgg$ for Majorana and real scalar pairs as a function of
    $r$ in the chiral limit $m_f = 0$ (the cross sections are given
    for $y_l= g_l = 1$ and $\mdm=100$ GeV).}
  \label{fig:svgg}
\end{figure}

Naively we would expect a similar result to hold for the annihilation
of scalar particles, with the proviso that the initial state is CP
even in this case, while it is CP odd for Majorana particles, thus by
replacing the chiral anomaly with the trace anomaly. Concretely, in
the limit $r = M_\psi/M_S \gg 1$, we would replace the heavy fermion
propagator in the box diagrams by the effective contact interaction of
(\ref{eq:effopT}) and then use the trace anomaly,
$\Theta^\mu_{\;\mu}~\propto~F_{\alpha\beta}F^{\alpha\beta}$ to
estimate the annihilation into gamma ray lines. The most immediate
consequence of this argument is that the annihilation amplitude should
be finite in the chiral limit, just like in the Majorana
case.\footnote{As a way of comparison, notice that this is very
  different from the contribution of light fermions to the decay of
  the Higgs in two photons, which vanishes in the chiral limit, see
  e.g.~\cite{Djouadi:2005gi}. } While this turns out to be the case,
the argument seems to be incorrect or, at the very least, it does not
give the dominant contribution to the process. Indeed the trace
anomaly would lead to a cross section that scales like $M_\psi^{-8}$
(the effective operator of (\ref{eq:effopT}) is dimension 8), while a
calculation of the full box diagram gives a results which actually
scales like $M_\psi^{-4}$, again as in the Majorana case. More
precisely, following the result quoted in~\cite{Tulin:2012uq}, which
is based on a calculation made in~\cite{Bertone:2009cb}, we get in the
chiral limit and for $r \gg 1$,
$$
\sigma v (S S \rightarrow \gamma \gamma) =  {\alpha^2\over 36 \pi^3} {y_l^4\over M_{\rm S}^2}{1\over r^4},
$$ which, all other things being kept constant, differs from the
corresponding expression in the Majorana case simply by a factor of
$16/9$. We are currently re-doing the calculations of the box diagram
made in~\cite{Bertone:2009cb}, for its own sake of and for the
possible implications for direct detection (see next section). For the
time being, we tentatively conclude that there is a subtle distinction
between the Majorana and scalar cases. This is further illustrated by
the fact that the exact dependence of $\sigma v (S S \rightarrow
\gamma \gamma)$ on $r$, which is given for reference in the appendix,
diverges at $r=1$ in the chiral limit, as illustrated in
Fig.~\ref{fig:svgg}.

\bigskip

\begin{figure}[t]
  \begin{tabular}{cc}
    \includegraphics[width=8cm]{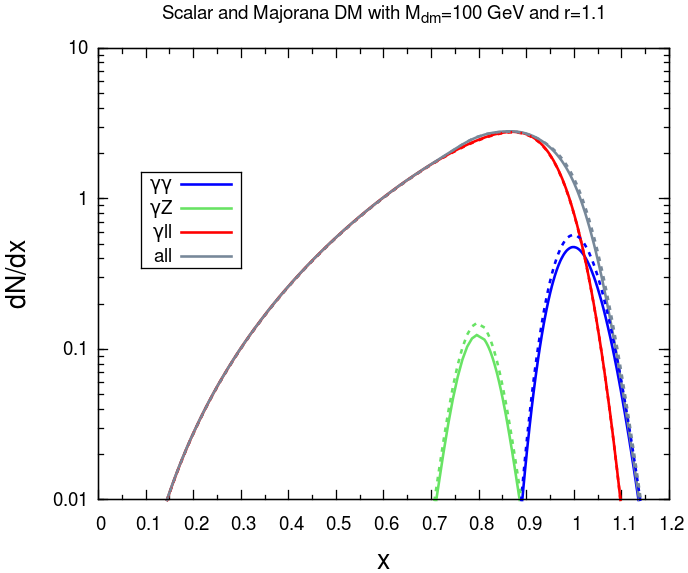}&
\includegraphics[width=8cm]{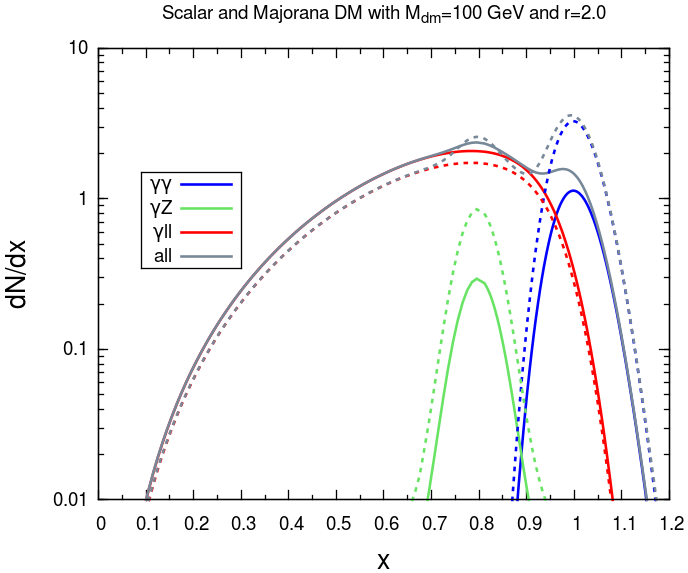}
  \end{tabular}
    \caption{Normalized spectra $\frac{dN_i}{dx}$ of
      Eq.~(\ref{eq:dNidx}) as a function of $x=E_\gamma/\mdm$ for two
      values of $r=1.1$ (left) and 2.0 (right) for scalar (continuous
      lines) and Majorana (dotted lines) dark matter with $\mdm=100$ GeV. The spectra have
      been convoluted with a gaussian kernel assuming a relative
      energy resolution $\Delta E/E=0.1$. }
  \label{fig:spectrtot}
\end{figure}

Using these results, we show as an illustration the differential
photon spectrum associated to the $\gamma\gamma$ (blue lines), $\gamma
Z$ (green lines) and $\gamma\bar l l$ (red lines) channels as well as
their sum (grey lines) in Fig.~\ref{fig:spectrtot}. The spectra have
been normalized by the sum of the annihilation cross-sections into the
three final states, i.e.
\begin{eqnarray}
  \frac{dN_i}{dx}&=&\frac{1}{\sveff_{\gamma}}\frac{d\sveff_i}{dx} \qquad\mbox{for }\quad i=\gamma\gamma,\gamma Z, \gamma\bar ll
\label{eq:dNidx}
\end{eqnarray}
where $x=E_\gamma/\mdm$ and $\sveff_{\gamma}=\sum_i\sveff_i$. We have chosen
$\mdm= 100$ GeV and we have considered annihilation into one single lepton
specie, with the Yukawa coupling set to one.
For these parameters, we have obtained the cross sections $\svgg$ and
$\svgll$ listed in Table~\ref{tab:m100}.  We have estimated the
$\svgZ$ cross-sections using
\begin{equation}
  \svgZ\simeq 2\,\tan[\theta_W]^2 \left(1 -
\left(\frac{M_Z}{2\mdm}\right)^2\right)^3\svgg\, ,
\end{equation}
 with $\theta_W$ is the Weinberg angle, as in the case of Higgs decay
 into two photons or $\gamma Z$ (see e.g.~\cite{Djouadi:2005gi}). From
 these numbers, we see once more that, for a fixed set of model
 parameters $\mdm$ , $y_l=g_l$ and $r$, the radiative
 processes are always larger for scalar particles, especially for Bremsstrahlung.

The photon raw spectra for the 3-body final state has been obtained in
Sec.~\ref{sec:spectr-cross-sect} while in the $\gamma\gamma$ case the
spectra is just a delta function at $E_\gamma=\mdm$, multiplied by two
to account for photon multiplicity. For the $\gamma Z$ final state,
one expects a broader feature, due to the Z width, around
$\mdm(1-M_Z^2/4\mdm^2)$.  We follow~\cite{Bertone:2009cb,Jackson:2009kg} that described the resulting
photon spectrum with an intrinsic width that depends on the final
state massive boson. All raw spectra are then convoluted with a
gaussian kernel in order to account for the finite energy resolution
of the detector. In Fig.~\ref{fig:spectrtot}, we consider a relative
energy resolution of $\Delta E/E=0.1$.

\begin{table}[t]
  \begin{tabular}{c|cc|cc}
     &Majorana& &Scalar&\\
    \hline
    &$r=$1.1&$r=2$&$r=1.1$&$r=2$\\
    \hline
$\svgll$&$1.2\, 10^{-27}$&$1.3\, 10^{-29}$&$10\, 10^{-27}$&$1\, 10^{-28}$\\
$\svgg$&$4.1\, 10^{-29}$&$3.08\,10^{-30}$& $2.7\, 10^{-28}$&$ 7.1\, 10^{-30}$\\
\hline
  \end{tabular}
  \caption{Cross-sections in units of cm$^3$/s for $\mdm=100$ GeV and
    $y_l=g_l=1$}
  \label{tab:m100}
\end{table}

From the normalized spectra presented in Fig.~\ref{fig:spectrtot} the
differences between Majorana and scalar dark matter are not
obvious. The general form of the spectra is actually very similar for
both scalar (continuous curves) and Majorana (dotted curves) particles
especially for $r\approx 1$ in which case the Bremsstrahlung drives
the main features of the full spectrum. For $r=2$, the scalar photon
line spectrum is still dominated by the VIB characteristics while in
the Majorana case a double line structure become more distinguishable
(in the Majorana case, we have obtained results which are consistent
with those of~\cite{Garny:2013ama}).  Notice that, for a more complete
description of the photon spectrum, one should also include the
continuous gamma spectrum from the $Z\gamma$ line as well as from the
production of the final state leptons, in particular in the case of
tau leptons. This however affects the spectra for small energies only,
and this only slightly, see for instance~\cite{Bringmann:2012vr}.

\section{Possible phenomenological implications}
\label{sec:pheno}

In this section, we first discuss the implication of d-wave
suppression of the annihilation cross section on the dark matter relic
abundance. This will lead us to the conclusion that larger Yukawas are
needed in the scalar dark matter case than in the Majorana one in
order to account for the total amount of dark matter. This in turn
will imply that the cross section for radiative 3-body processes can
become as important as the 2-body process in the scalar case. Finally
in order to illustrate our findings we present a numerical analysis
comparing the viable parameter space for the simple scalar and
Majorana dark matter models defined in Sec.~\ref{sec:model}.

\subsection{Relic abundance for d-wave suppressed annihilation}
\label{sec:relic-abundance}

To begin with, we consider the simplest scenario and assume that the
relic abundance of pair of scalar in the early universe is determined
by the d-wave suppressed 2-body process discussed in Sec.~\ref{sec:model}. The thermally averaged cross section is given by
$$
\langle \sigma v\rangle = \sum_f { y_l^4 \over 60 \pi}{\langle v^4\rangle \over M_S^2}{1\over (1+r^2)^4}.
$$
Following \cite{Srednicki:1988ce} (or \cite{Gondolo:1990dk}) and using
$$
\langle \sigma v \rangle = {\int d^3 k_1 d^3 k_2 e^{-(E_1 + E_2)/T} \sigma v\over \int d^3k_1 d^3k_2 e^{-(E_1+E_2)/T}} = \sqrt{x^3\over 4 \pi} \int_0^\infty dv \, e^{- x v^2/4} \sigma v
$$
with $x = M_{\rm dm}/T$, we get
$$
\langle v^2\rangle = {6\over x_f} \quad {\rm and} \quad \langle v^4\rangle = {60\over x_f^2}\,.
$$ Using $x_f = 25$ for the relative temperature at freeze-out
one gets $\langle v^2\rangle= 0.24$ and $\langle v^4\rangle=
0.1$. One has also to take into account the fact that the velocity has
a slight numerical impact on the $x_f$ dependence of the relic
abundance. If the thermal averaged cross section scale like $x^{-n}$,
with $n=0,1,..$ for s-wave, p-wave,... dominated cross section, taking
into account that the Boltzmann equation for the relic abundance for
$x > x_f$ takes the form
$$
{dY_{\rm dm}\over dx} = - {\lambda\over x^{2+n}} Y_{\rm dm}^2
$$ where $\lambda$ is a constant, and $Y_{\rm dm}$ denote the comoving
dark matter number density then \cite{Kolb:1990vq}
$$
Y_{\rm dm}^\infty \approx {(n+1) x_f^{n+1}\over \lambda}.
$$ Hence, to reach the same relic abundance $\Omega_{dm}h^2$, the
averaged annihilation cross section $\langle \sigma v \rangle$ in
for instance a pure d-wave channel must be larger by a factor of $n+1=3$
wrt the s-wave case i.e. $\langle \sigma v \rangle_{\rm d-wave}
\approx 9 \cdot 10^{-26}\, \cm^2\s^{-1}$.  Considering the velocity
expansion of the annihilation cross section to the next order compared
to e.g.~\cite{Griest:1990kh,Kolb:1990vq}, writing $\sv=a+bv^2+cv^4$,
we have
\begin{equation}
  \Omega_{dm}h^2\simeq\frac{1.07\, 10^9 \, x_f}{M_{pl}/{\rm GeV}\sqrt{g_*} (a+3b/x_f+20 c/x_f^2)}
\end{equation}
where $M_{pl}= 1.22 \, 10^9 $ GeV is the Planck mass and $g_*$ is the
number of relativistic degrees of freedom at the time of freeze-out.

Lest the reader think that we are splitting hairs, consider again the ratio of (\ref{eq:ratio}) but now expressed in terms of the relic abundances, 
$$ 
{\Omega_{dm}\vert_S\over 
\Omega_{dm}\vert_\chi} = {\sum_l g^{ 4}_l\,  (5 \times 3)\, (1+r^4)\, x_f^2 \over \sum_l y_l^4 \, (4\times 20)\, x_f} = \left({\sum_l g^{ 4}_l \over \sum_l y_l^4}\right)\; {3\over 16}  x_f (1+r^4)\,.
$$
 The factor that multiplies the couplings is typically ${\cal O}(10)$, hence larger Yukawa couplings are required to reach same abundance for the scalar than for the Majorana. It is pretty clear that this implies larger Bremsstrahlung emission in the case of scalar DM. 

\subsection{Enhanced three-body processes for  scalar dark matter}
\label{sec:enhanced-three-body}
\begin{figure}[t] 
\begin{center}
  \begin{tabular}{cc}
    \includegraphics[width=8cm]{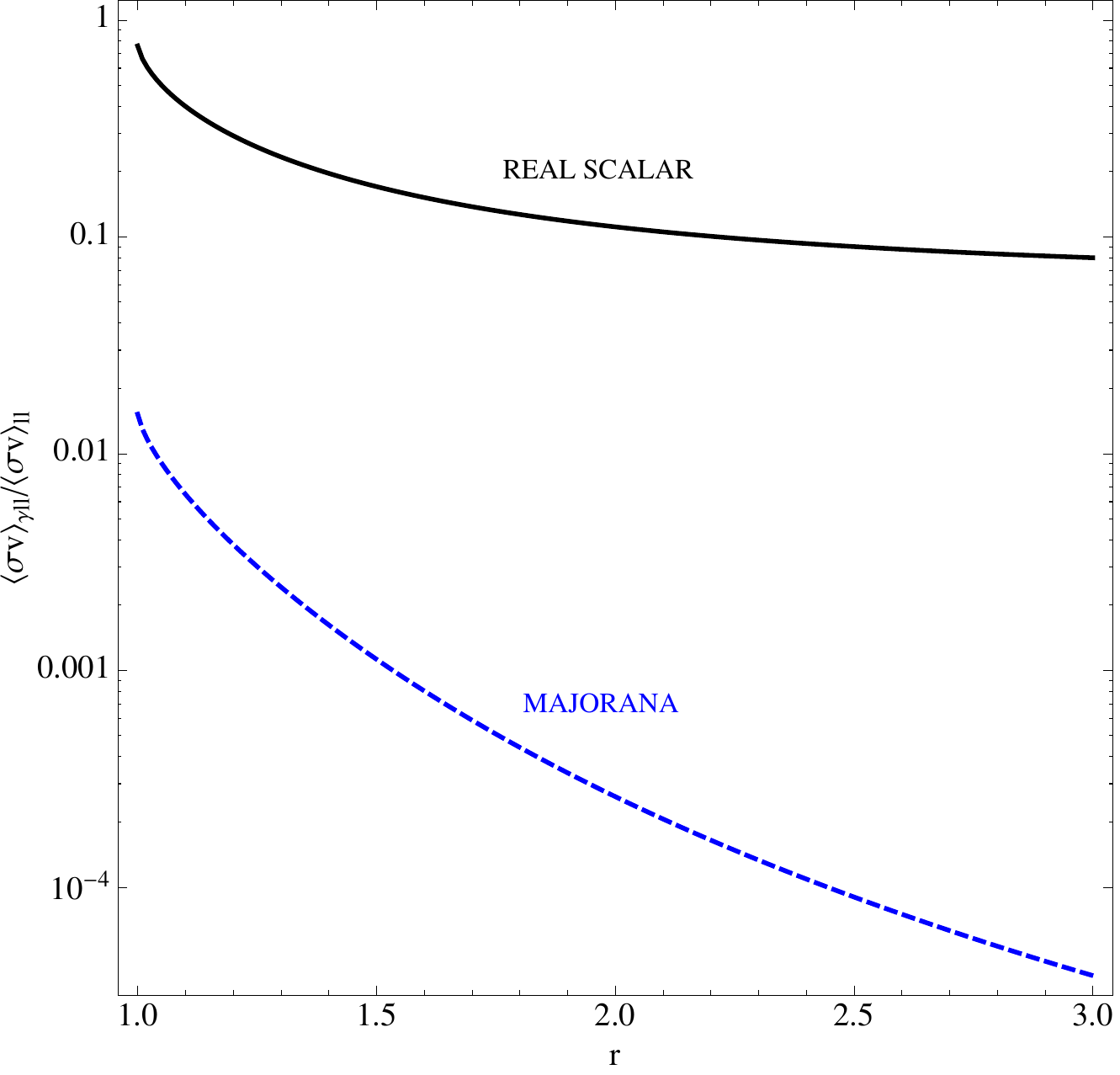}
   &\includegraphics[width=8cm]{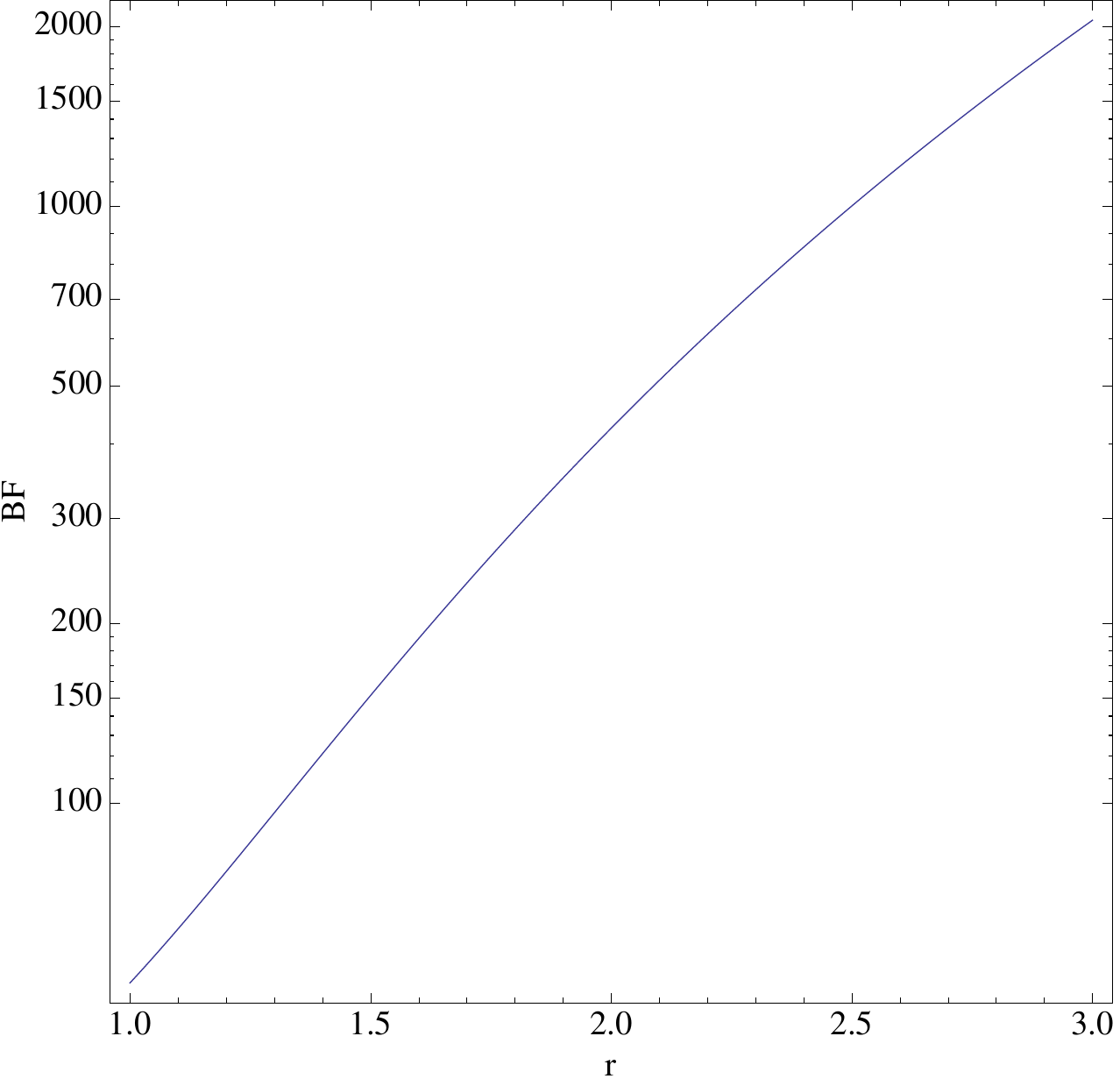}
  \end{tabular}
  \end{center}
\caption{Left: Ratios of 3-body to 2-body total annihilation cross
  section at the time of freeze-out for Majorana (blue dashed line)
  and scalar (black continuous line) DM as a function of the mass
  ratio $r$. Right: Maximal relative enhancement (boost factor $BF$ of
  Eq.~(\ref{eq:boost})) of the Bremsstrahlung signal from the
  annihilation of scalar DM compared to the Majorana case (see
  text).  }
\label{fig:ratios}
\end{figure}
In the Majorana case, the 3-body annihilation cross section is always
small compared to the 2-body one, which is relevant for the abundance
in the early universe. The ratios ${\svgll/ \svll}$ are shown in the
left hand side of Fig.~\ref{fig:ratios} for Majorana (blue dashed
line) and scalar (black continuous line) dark matter. The limiting
values for $r\rightarrow 1$ are 
\begin{equation}
\left. {\svgll\over \svll}\right\vert_\chi \approx 0.015 \qquad \mbox{and}\qquad
\left. {\svgll\over \svll}\right\vert_S \approx 0.76 \label{eq:ratio_s}
\end{equation}
taking $x_f = 25$. 
It is by itself remarkable that, for the scalar candidate, the radiative process may be almost as important as the 2-body one. We will come back to this in the next section. In the meantime we may define the ``boost factor'' ($BF$)
\begin{equation}
   BF = \left. {\svgll\over \svll}\right\vert_S  \times \left. {\svll\over \svgll}\right\vert_\chi\, ,
\label{eq:boost}
\end{equation}
that is equal to 50 for $r\rightarrow 1$ and gives the relative
enhancement of the Bremsstrahlung signal of the scalar DM candidate
compared to the Majorana one. It should be clear from the behaviour of
the cross section that $BF=50$ is actually a minimum and this is further
illustrated in the right hand side of Fig.~\ref{fig:ratios}.

The enhancement of the Bremsstrahlung signal from a scalar WIMP is our
main result, but as such it is of no immediate use, as other
processes may determine the relic abundance. In particular one has to
take into account co-annihilation processes, which are important in
the case of nearly degenerate particles, $r \lesssim 1.1$
\cite{Griest:1990kh}. Also a singlet scalar candidate may have
renormalizable coupling to the SM scalar (SMS).
To study these effects, we have implemented the scalar and Majorana
models in Micromegas~\cite{Belanger:2010gh} with the help of
Feynrules~\cite{Christensen:2008py}. Our analysis is detailed in the
next section.

\subsection{Numerical analysis} 
\label{sec:numerical-analysis}
\begin{figure}[h!]
  \begin{center}
    \hspace{-1cm}
    \begin{tabular}{cc}
    \includegraphics[width=8.5cm]{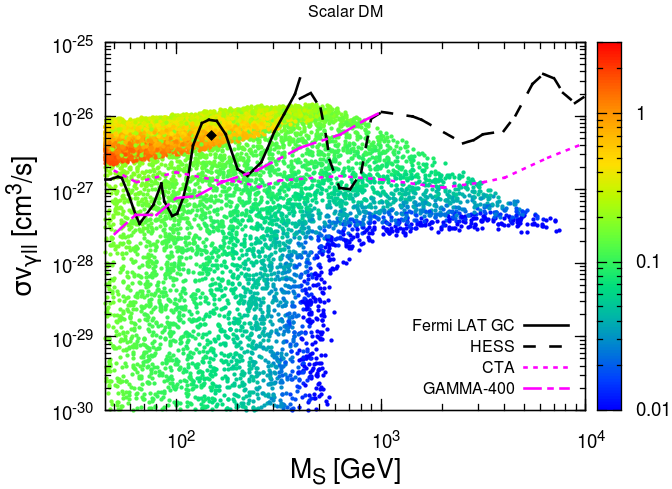} &
    \includegraphics[width=8.5cm]{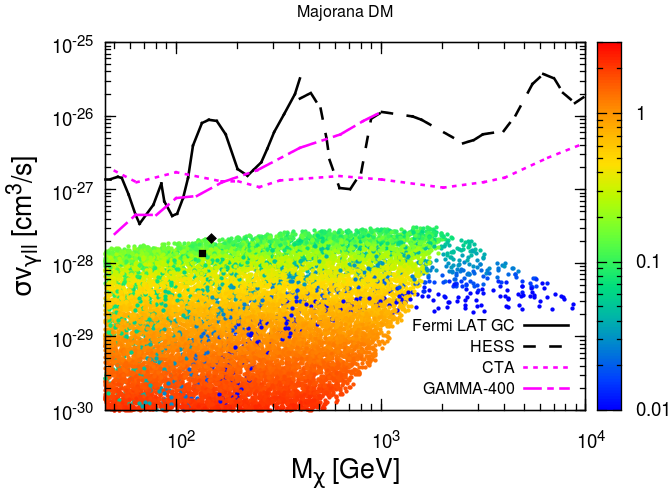}
    \end{tabular}
  \end{center}
  \caption{ Annihilation cross section into three-body final states
    $\svgll$ for WIMP candidates (left: scalar, right: Majorana
    fermion) coupling to one single family of massless leptons. The
    lines appearing in the upper part of the plots give an upper
    bounds on $\svgll$ for $r=1.1$ associated to Fermi LAT, HESS
    experiments as well as future constraints from CTA and
    GAMMA-400~\cite{Garny:2013ama}.  The filled diamond correspond to
    the benchmark models of Tab.~\ref{tab:m149} and the filled
    rectangle corresponds to the benchmark studied
    in~\cite{Bergstrom:2012bd}. The color gradient scale is associated
    to the values of $r-1$.}
  \label{fig:sv3}
\end{figure}

\begin{figure}[!]
  \begin{center}
    \begin{tabular}{cc}
    \includegraphics[width=8.5cm]{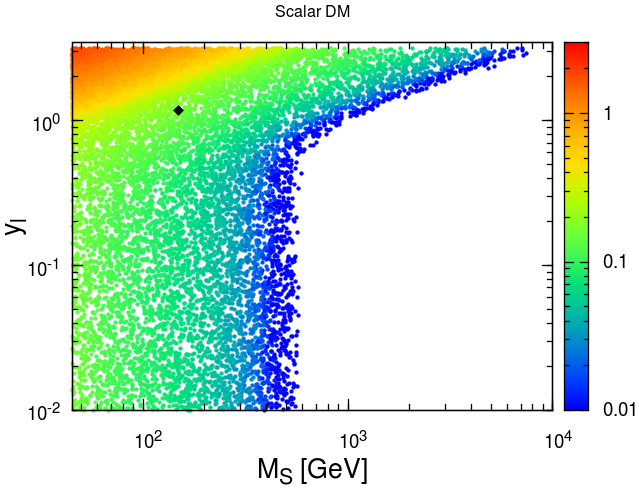} &
    \includegraphics[width=8.5cm]{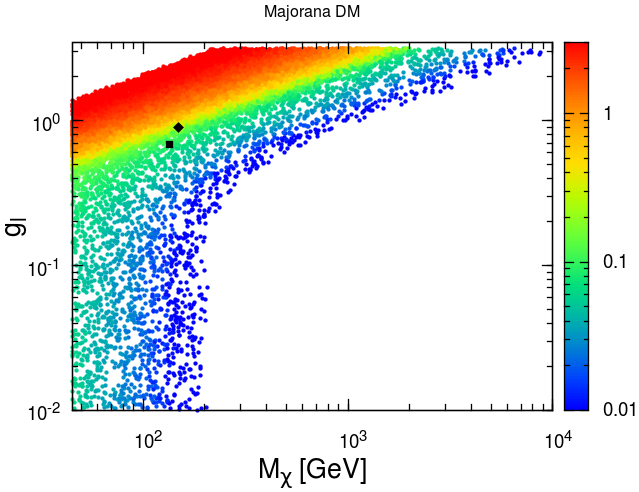}
    \end{tabular}
  \end{center}
  \caption{Yukawa coupling $y_l$ and $g_l$ needed to get the observed
    relic abundance (left: scalar, right: Majorana fermion). The
    filled diamond corresponds to the benchmark models of
    Tab.~\ref{tab:m149} and the filled rectangle correspond to the
    benchmark studied in~\cite{Bergstrom:2012bd}. The color gradient
    scale is associated  to the values of $r-1$.}
  \label{fig:yuk}
\end{figure}

 We have considered the scalar and Majorana dark matter models which
 interaction with the SM fermions is dictated by Eqs.~(\ref{eq:yuk1}) and~(\ref{eq:majDM}), neglecting extra interactions through the SM
 scalar portal from Eq.~(\ref{eq:ls}), i.e. setting $\lambda_S=0$. In
 addition, we have assumed that the dark matter couples to one single
 lepton specie,  the electron. The result of a random scan over the
 parameter space is shown in Figs~\ref{fig:sv3} and \ref{fig:yuk}, see
 also the appendix for more details. We also give the 3-body
 annihilation cross section (Yukawa coupling) versus the mass of the
 DM in Fig.~\ref{fig:sv3} (resp. Fig.~\ref{fig:yuk}) both for the scalar
 (left) and Majorana (right) candidates. All the points match the
 observed cosmological relic abundance. Notice that we have taken into
 account the bremsstrahlung contribution to the effective annihilation
 cross section relevant for the computation of the relic abundance. In
 the case of the scalar dark matter such process can modify the dark
 matter abundance up to a 15\% for the largest values of the Yukawa
 couplings while in the case of the Majorana dark matter it may be
 safely neglected (it is always $<0.5$\%).

The color code corresponds to different values of $r-1$, which is the relative mass difference between the DM particle and the heavy charged particle. The blueish
points (roughly the lower right points in the plots) correspond to
dark matter candidates which mass is nearly degenerate, roughly $r\sim
1$, with the mass of the electrically charged heavy particle ($\Phi$ and
$\Psi$). In the latter case, the annihilation of the charged particles 
\begin{eqnarray}
  \bar \Psi\Psi \quad \mbox{or}\quad \Phi^\dag\Phi&\rightarrow& \bar q q,
  \bar l l, \gamma\gamma
\label{eq:annNLZP}
 \end{eqnarray}
 and the co-annihilation processes 
\begin{eqnarray}
  \bar \Psi S \quad \mbox{or}\quad \Phi^\dag\chi&\rightarrow& \bar e \gamma
\label{eq:coann}
 \end{eqnarray}
 are important for the determination of the relic abundance.  The
 greenish points (roughly the top regions) correspond to candidates
 for which the dark matter annihilation into lepton-antilepton pairs,
 \begin{equation}
   SS \quad \mbox{or}\quad \chi\chi\rightarrow \bar e e
\label{eq:ann}
 \end{equation}
become progressively more important and so does Bremsstrahlung.
 Increasing further the ratio of masses $r$, the relative weight of
the 3-body process diminishes, particularly for the Majorana case, see
Sec.~\ref{sec:spectr-cross-sect}.
The maximal value of $r$ is reached for $M_S \approx 400$ GeV and
$M_\chi \approx 1$ TeV given our assumption on the Yukawas $y_l,
g_l<\pi$, see Fig.~\ref{fig:yuk}. In our numerical analysis as well as
writing the co-annihilation annihilation processes as in
(\ref{eq:coann}) and~(\ref{eq:ann}), we have assumed that the coupling
of Eqs.~(\ref{eq:yuk1}) and (\ref{eq:majDM}) is made for $l\equiv
e$. For dark matter coupling to three light flavours, the results of
Fig.~\ref{fig:yuk} should be rescaled by a factor of $
3^{-1/4}\approx 0.76 $. From Fig.~\ref{fig:yuk}, it is clear that significant
Bremsstrahlung requires rather large Yukawa couplings $y_l,g_l \sim
1$, especially for scalar candidates. This simply reflects the fact
that the annihilation is d-wave for scalars and p-wave for Majorana,
as discussed in Sec.~\ref{sec:model}.

For a fixed value of $r$ and $r\approx 1$, one clearly distinguish two
different regimes in the Yukawa-$\mdm$ plane. For the lowest values of
the Yukawas, the dark matter models accounting for dark matter
abundance are clearly independent of $y_l$ or $g_l$. At some point the
Yukawa function of $\mdm$ begins to bend and finally reaches a regime
with $\log(y_l,g_l) \propto \log(\mdm)$. This behaviour can be
understood by comparing the dependence of the processes
(\ref{eq:annNLZP})-(\ref{eq:ann}) in $y_l$ or $g_l$. For the smallest
values of the coupling and $r-1$, it is the annihilation processes of
$\Psi$ and $\chi$ through $\gamma$ and $Z$ boson exchanges
in~(\ref{eq:annNLZP}) that dominate over all other processes and play
the major role in fixing the dark matter abundance.  The corresponding
cross section is $\propto g^4$, with $g$ the weak coupling, and it is
independent of $y_l,g_l$ (see Appendix).  Thus no Yukawa or $\mdm$
dependence is to be expected in this regime. For larger values of the
Yukawas, the coannihilation cross section of~(\ref{eq:coann}) $\propto
g^2y_l^2$ or $ g^2g_l^2 $  begin to
compete with charged particle annihilation and the abundance begins to
depend on the Yukawa coupling. The relative importance of both
of those regimes is weighted by Boltzmann factors, $\exp[-(r-1)\,x_f]$
and $\exp[-(r-1)\,x_f]^2$ for (\ref{eq:coann}) and (\ref{eq:annNLZP})
respectively~\cite{Griest:1990kh}, so that the dependence in $y_l,g_l$
becomes more pronounced for larger values of $r$. For the largest
values of the Yukawas, the processes which are $\propto y_l^4$ or
$g^4_l$ fix the dark matter density. One should be aware that the
standard treatment of freeze-out mechanism and coannihilation
processes~\cite{Griest:1990kh}, as implemented in numerical code like
Micromegas~\cite{Belanger:2010gh}, rests on the assumption that the
dark matter and heavy charged particles are in chemical
equilibrium. Here, as in~\cite{Garny:2013ama}, we have simply checked
under which conditions the processes $\chi \;l\leftrightarrow \Phi
\;\gamma$ and $S \;l\leftrightarrow \Psi\; \gamma$ are in equilibrium
at the epoch of thermal freeze-out: this should be the case provided
$y_l, g_l>10^{-3}$. 

Taken the Bremsstrahlung spectral features seriously, one may wonder
if these candidates are compatible with constraints from the current
gamma ray experiments, in particular Fermi-LAT and HESS. To this end,
we also report in Fig~\ref{fig:sv3}, with a series of upper bounds on
$\sigma_{\gamma ll }+2\sigma_{\gamma\gamma}$ for $r=1.1$ that
were derived in~\cite{Garny:2013ama} using Fermi-LAT and HESS data as
well as the future constraints from the GAMMA-400 satellite
mission and the Cerenkov Telescope Array (CTA). For simplicity, we
report the limits in the $\mdm-\svgll$ plane without
adding the contribution from $\sigma_{\gamma\gamma}$, which are anyway
negligible for $r \lesssim 2$, see Sec.~\ref{sec:annih-into-gamma} and
also~\cite{Garny:2013ama}.  As in~\cite{Bringmann:2012vr}, we
find that the largest possible value of the 3-body cross section for a
Majorana dark matter giving rise to right dark matter abundance is
always substantially smaller than current limits (so that an
astrophysical boost would be required to match any excess, say the
possible feature around 130).  In contrast, in the scalar dark
matter case, within the assumptions made so far regarding e.g.
$\lambda_S=0$, one can easily cross those limits, typically giving
rise to a more important gamma-ray flux.
\begin{table}[ht]
  \begin{tabular}{c|cc|cc|cccc}
    Benchmarks& $y_i$ &$r$&$\svgll$&$\svgg$&$\Omega_{\rm dm} h^2$&$R_{\rm 3bdy}$&$R_{\rm ann}$&$R_{\rm co}$\\
    \hline
    Scalar& $y_l=1.17$&1.16&$5.4\, 10^{-27}$& $1.3\,10^{-28}$&0.11&0.06&0.28&0.41\\
    Majorana&$g_l = 0.9$&1.17&$2.2\, 10^{-28}$& $8.9\,10^{-30}$&0.10&0.002&0.95&0.047\\
    \hline
  \end{tabular}
  \caption{Benchmark models for dark matter candidates with
    $\mdm=150$ GeV which VIB signal could be associated to a gamma ray
    excess around 130 GeV. Cross sections are given in units of
    cm$^3$/s, $y_i$ refers to the Yukawa couplings $y_l$ and $g_l$ and
    $R_{\rm 3bdy}$, $R_{\rm ann}$ and $R_{\rm co}$ give the relative
    contribution of 3-body, annihilation and coannihilation processes,
    respectively, effectively contributing to the relic abundance. }
  \label{tab:m149}
\end{table}

For the sake of illustration we consider two benchmark models (see
Table \ref{tab:m149}) that could be relevant for the possible excess
of gamma rays around $E_\gamma=$130 GeV in the Fermi-LAT
data~\cite{Bringmann:2012vr,Weniger:2012tx}. Both candidates have a
relic abundance $\Omega_{\rm dm }h^2\sim 0.1$. The relative
contributions of the various processes to their annihilation cross
section at freeze-out~\cite{Griest:1990kh} are given by $R_{\rm 3bdy}$
for dark matter annihilation into the 3-body channels, $R_{\rm ann}$
for annihilation into 2-body and $R_{\rm co}$ for coannihilation
processes, while the annihilation of the heavy charged particles
contribute for $1-R_{\rm 3bdy}-R_{\rm ann}-R_{\rm co}$.  The dark
matter mass is taken to be 150 GeV, following~\cite{Bringmann:2012vr}
that pointed out that the best-fit 3-body cross section is
$\sveff_{\rm 3bdy}^{\rm best}\sim 6.2 \,10^{-27}$ cm$^3$/s with $\mdm
\sim 150$ GeV --- in the case of a gamma ray line, the best fit is for
about $\sveff_{\gamma\gamma} \approx 1.27 \,10^{-27}$ cm$^3$/s with a
mass of $\sim 130$ GeV. From Fig.~\ref{fig:sv3}, it is clear that
$\sveff_{\rm 3bdy}^{\rm best}$ can be reached for a scalar dark matter
candidate, while an extra, possibly astrophysical boost factor would
be required for a Majorana candidate.

\begin{figure}[t]
  \begin{center}
    \includegraphics[width=10cm]{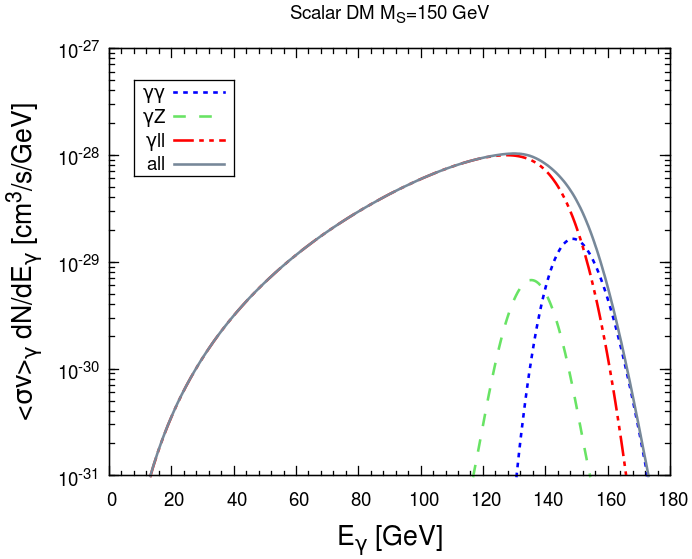}
   \end{center}
    \caption{Spectra $\sveff_\gamma\frac{dN_i}{dx}$, with
      $\sveff_\gamma$ defined as in Eq.~(\ref{eq:dNidx}), as a
      function of $E_\gamma$ for the scalar dark matter benchmark
      model of Table~\ref{tab:m149} ($\mdm=150$ GeV). The spectra have
      been convoluted with a gaussian kernel assuming a relative
      energy resolution of $\Delta E/E=0.1$. }
  \label{fig:spectr149}
\end{figure}

In Table~\ref{tab:m149}, the Majorana benchmark candidate is chosen so
as to maximize the 3-body annihilation cross section.  In order for
this candidate to saturate the data, a boost factor of about 10 would
be needed.  The scalar benchmark has also $\mdm \sim 150$ GeV but this
time the Yukawa coupling is chosen so that the candidate could
account
for a gamma ray at $E_\gamma
\sim 130$ GeV, see the resulting spectra in Fig.~\ref{fig:spectr149}.
Unfortunately no distinction can be made between the shapes of scalar
and Majorana benchmark spectra, even for optimistic resolutions such
as $\Delta E/E\sim 0.02$, we thus only show the scalar dark matter
one. The points associated to our benchmark models correspond to the
filled black diamonds in Figs.~\ref{fig:sv3} and \ref{fig:yuk}. For
reference, we also report the $\mdm \sim 130$ GeV candidate put
forward in \cite{Bergstrom:2012bd}, which is assumed to couple to the
three families of leptons with universal Yukawa coupling $g_l = 0.52$
for $l=e,\mu, \tau$ and to a heavy scalar ($\SH$) with mass such that
$r = M_{\SH}/M_\chi = 1.1$.
\footnote{Notice that the value of $\svgll=1.2 \, 10^{-28}$
  obtained here is two times larger than the number reported
  in~\cite{Bergstrom:2012bd}. We expect that this is due to a factor
  of two rescaling of $\svgll$ in~\cite{Bergstrom:2012bd} which is
  relevant when compared to gamma ray flux constraints on $\svgg$.}

\subsection{Prospects and possible extensions}
\label{sec:prospects}

So far we have considered a specific scenario, assuming the coupling
of a real scalar DM candidate to a heavy vector-like leptonic particle
which we take to be singlet under $SU(2)$. This we have done for the
sake of simplicity and illustration of the possible enhancement of the
VIB in the case of a scalar compared to a Majorana DM candidate. A
systematic analysis of the possible extensions of this scenario is
beyond the scope of the present work, and will be presented
elsewhere. In this section we briefly summarize some possible
forthcoming results.

One possible improvement would be to consider the continuous gamma ray
emission associated to weak gauge bosons (which we have neglected
here) especially generalizing the interaction~(\ref{eq:yuk1}) to the
case of Yukawa couplings to $SU(2)$ doublet vector-like leptons. 
In the latter case, new contribution $SS\rightarrow W\bar l \nu$ to
3-body processes should be taken into account.\footnote{ Notice that
  in our numerical analysis we took into account the annihilation into
  $Z \bar l l$ for the calculation of the relic abundance which
  typically give rise to $\sveff_{Zll}\sim \svgll/3$} Beside the
obvious boost of the signal that we may expect in the scalar case
compared to the Majorana case (which would couple to new a scalar
doublet), the results will however match the existing analysis
of~\cite{Garny:2011ii}, since the scalar and Majorana dark matter
spectra are expected to be precisely identical.

Possibly more interesting would be to consider Yukawa couplings to
heavy vector-like quarks. In this case, there are potentially two
aspects that are worth being studied in more details. To begin with,
we have the possibility of the VIB of a gluon. As emphasized in the
previous section, in the case of scalar particles annihilation, the
VIB of a photon is substantial, possibly similar in magnitude with the
2-body process, at least for annihilation in the early universe. The
ratio of the VIB of a gluon to that of a photon being given by \be
\label{eq:ratiogluegamma}
{\langle \sigma v \rangle_{g \bar q q}\over \langle \sigma
  v\rangle_{\gamma \bar q q}} = {N_c^2 -1\over 2 N_c}\, {\alpha_s\over
  Q^2\alpha} \ee which is about $\sim 40$ for up-like quarks, and
$\sim 150$ for down-like quarks. These features imply that the VBI of
a gluon will be potentially more relevant in determining the relic
abundance than the corresponding two body process, at least for light
quarks. This in turn would imply that the annihilation rate into gamma
rays is fixed by (\ref{eq:ratiogluegamma}) a possibility very much in
the spirit of the scenario considered in~\cite{Chu:2012qy}. We will
address such processes with gluons and quarks in final states in a
work in progress, as well as the constraints from the measurements of
the antiproton flux in cosmic rays, see
e.g.~\cite{Ibarra:2012dw,Chu:2012qy}.

A second interesting aspect in scalar dark matter models coupled to
heavy vector-like quarks is direct detection.  A priori the analysis
should be similar to the Majorana case discussed
in~\cite{Ibarra:2012dw}. In particular, for the scalar effective
coupling to quarks, the relevant interaction should be related to the
effective operator (\ref{eq:effopS}), with the obvious substitution of
leptons by quarks. We expect though some subtle differences for the
contribution of the so-called twist-2 operator, which is formally
related to the operator (\ref{eq:effopT}). As is well known, in the
Majorana case the proper determination of direct detection collision
cross section requires the evaluation of the box diagram in kinematic
regimes in which the direct use of the trace anomaly may give wrong
results \cite{Drees:1993bu}. The equivalent process for scalar DM
through a vector-like portal has not yet been studied and will be
addressed in a future work.

 In our work, we have assumed that the quartic coupling of the real
 scalar to the Standard Model scalar may be neglected. An obvious
 generalization of the results presented here would be to relax such
 an assumption. For the time being, we just emphasize the trivial fact
 that this coupling may help decreasing the VIB cross section. Indeed,
 as shown in Fig.~\ref{fig:sv3}, there are many WIMP scalar candidates
 with a too large flux into gamma rays, and this all the way up to
 $\mdm \sim 900$ GeV. If another annihilation channel is opened, like
 through the SM scalar, the Yukawa coupling may be smaller, and these
 candidates may become viable. This is not quite the same for the
 Majorana candidates, for which the signal is systematically below the
 current constraints, and thus which would require some astrophysical
 boost to saturate the observations. As an
 example, let us take from our scans a scalar candidate with
 $\mdm=150$ GeV, $r=1.32$ and $y_l=1.8$ which relic abundance
 $\Omega_{\rm dm }h^2=0.1$ but, $\svgll=1.1\, 10^{-26}$ cm$^3$/s is
 excluded by Fermi-LAT constraint. By allowing for $\lambda_S=-0.06$
 and decreasing $y_l$ to 1.5, we can still account for $\Omega_{\rm dm
 }h^2=0.1$ while getting a $\svgll=5.3 \, 10^{-27}$ cm$^2$/s below the
 limits and near $\svgll^{\rm best}$.  Notice though that in the
 latter case, new channels $SS\rightarrow WW, ZZ, HH$ open and give
 rise to a continuous gamma ray component with a 2-body annihilation
 cross section that is no more d-wave suppressed. In addition, direct
 detection searches can also test such a scenario given that
 the scattering cross-section on a proton amounts to $8.45\, 10^{-46}$
 cm$^2$ and is nearly excluded by the Xenon100
 limits~\cite{Aprile:2012nq}. See also~\cite{Cline:2013gha} for recent
 study of latest and future constraints on singlet scalar dark matter
 and SMS portal.

In a broader perspective, one may consider a much more extended parity
odd sector. For definiteness, consider the so-called Inert Doublet
Model (IDM)~\cite{Deshpande:1977rw,Ma:2006km,Barbieri:2006dq}. In its
simplest incarnation it consists of the addition of a single scalar
doublet, $H_2$, odd under a $Z_2$ parity, with no expectation value.
A simple and very interesting extension consists in the addition of an
odd right-handed neutrino field \cite{Ma:2006km} with a Majorana mass
term, a model in which the mass of the SM neutrinos is generated
radiatively at one loop. In the same spirit, we may consider the
possibility of introducing heavy vector-like doublets, be them lepton
or quark-like, all odds under $Z_2$. Of course this scenario implies
the introduction of a large number of new parameters, to begin with
the Yukawa couplings, but it also broadens the range of possibilities
for the IDM, in particular regarding not only gamma ray
features,\footnote{Notice that the vanilla IDM already presents VBI
  gamma ray features \cite{Garcia-Cely:2013zga}.} but also potentially
antimatter in cosmic rays, new direct detection channels, and particle
physics signatures with flavour changing processes such as $\mu
\rightarrow e + \gamma$.~\footnote{Notice that such processes are
  potentially present in the singlet scalar scenario discussed here
  but are tuned to be irrelevant by taking the matrix Yukawa
  couplings, to be diagonal in lepton flavour space. See for
  instance~\cite{Cheung:2004xm} for an analysis in Majorana dark
  matter case and~\cite{Ishiwata:2013gma} for a recent study of flavour
  changing process in the vector-like portal scenario. }

\section{Conclusions}

In this work we have discussed a simple dark matter model that may
lead to a significant gamma ray spectral feature. It consists of real
scalar DM particle that interacts with SM leptons through heavy
vector-like charged fermions (the so-called vector-like portal). The
most striking feature of this model is the possibility of an enhanced
annihilation of dark matter in a process with virtual internal
bremsstrahlung (VIB). This rest on the fact that, 1/ in the chiral
limit the annihilation cross section in lepton antilepton pairs is
d-wave suppressed, i.e. $\propto v^4$ and, taking all other things
constants (dark matter mass, mass of the intermediate particle and the
Yukawa couplings) 2/ the bremsstrahlung cross section is relatively
larger than in the case of annihilation of Majorana particles. These
two features taken together imply that the VIB feature is much more
enhanced for real scalar dark matter compared to Majorana
candidates. 

For the sake of illustration, we have studied in more details a
concrete, albeit simplistic case of a real scalar with a Yukawa
coupling to the SM right-handed electron (or equivalently for all
practical purpose with universal coupling to all lepton). Our main
result may be read from Fig.~\ref{fig:sv3} that shows the annihilation
cross section into 3--body final state $\svgll$, and thus the associated
gamma ray flux, is comparatively much stronger for scalar
than for Majorana dark matter, provided that these candidates account for all the dark matter cosmological abundance.
For completeness, we have also compared the annihilation rates in
monochromatic gamma rays of real scalar and Majorana candidates.  In
the near future, we intend to generalize this framework to the case of
heavy vector-like quarks.

\vspace{5mm}
\section*{Acknowledgement}
\vspace{-3mm} 

We acknowledges stimulating discussions with Thomas Hambye and
C\'eline Boehm. F.G and M.T. work is partially supported by the
FNRS-FRS and the IISN.  LLH is supported by an ``FWO Vlaanderen"
post-doctoral fellowship project number 1271513 and recognizes partial
from the Strategic Research Program ``High Energy Physics" of the
Vrije Universiteit Brussel.  All the authors acknowledge support from
the Belgian Federal Science Policy through the Interuniversity
Attraction Pole P7/37.  \\ {\bf Note added :} while our work was
being completed, the article~\cite{Toma:2013bka} was published on the
ArXiv, finding similar results for real scalar dark matter.  \appendix

\section*{Appendix}
\label{sec:Xsections}
In this appendix, we provide some more expressions of annihilation
cross-sections involved in the determination of the relic abundance or
the gamma ray signal. Notice though that our numerical results of
Sec.~\ref{sec:numerical-analysis} were obtained using Micromegas code
which automatically compute all the two body processes at tree level
so that their expressions given below in the non-relativistic limit
are only given for the sake of completeness. We implemented the dark
matter 3-body annihilation cross-sections into $\gamma \bar l l$ and
$Z \bar l l $ in Micromegas. Also, notice that annihilation into two
gammas were neglected in the computation of the relic abundance.

In Sec.~\ref{sec:pheno}, we neglected s- and p-wave contributions to
the scalar dark matter annihilation cross-section. Here we give
the s- to d-wave contributions to the 2-body annihilation cross
section $S S \rightarrow l\bar l$, to leading order in $m_f$ and $v$,
\begin{equation}
\sigma v (S S \rightarrow \bar l l) =\frac{y^4_f}{4\pi}\frac{1}{M^2_S(1+r^2)^2}\left[\frac{m^2_f}{M^2_S}-\frac{4}{6}\frac{m^2_f v^2(1+2 r^2)}{M^2_S(1+r^2)^2}+\frac{1}{15}\frac{v^4}{(1+r^2)^2}\right].
\end{equation}

\begin{figure}[t]
\begin{center}
\includegraphics[width=8cm]{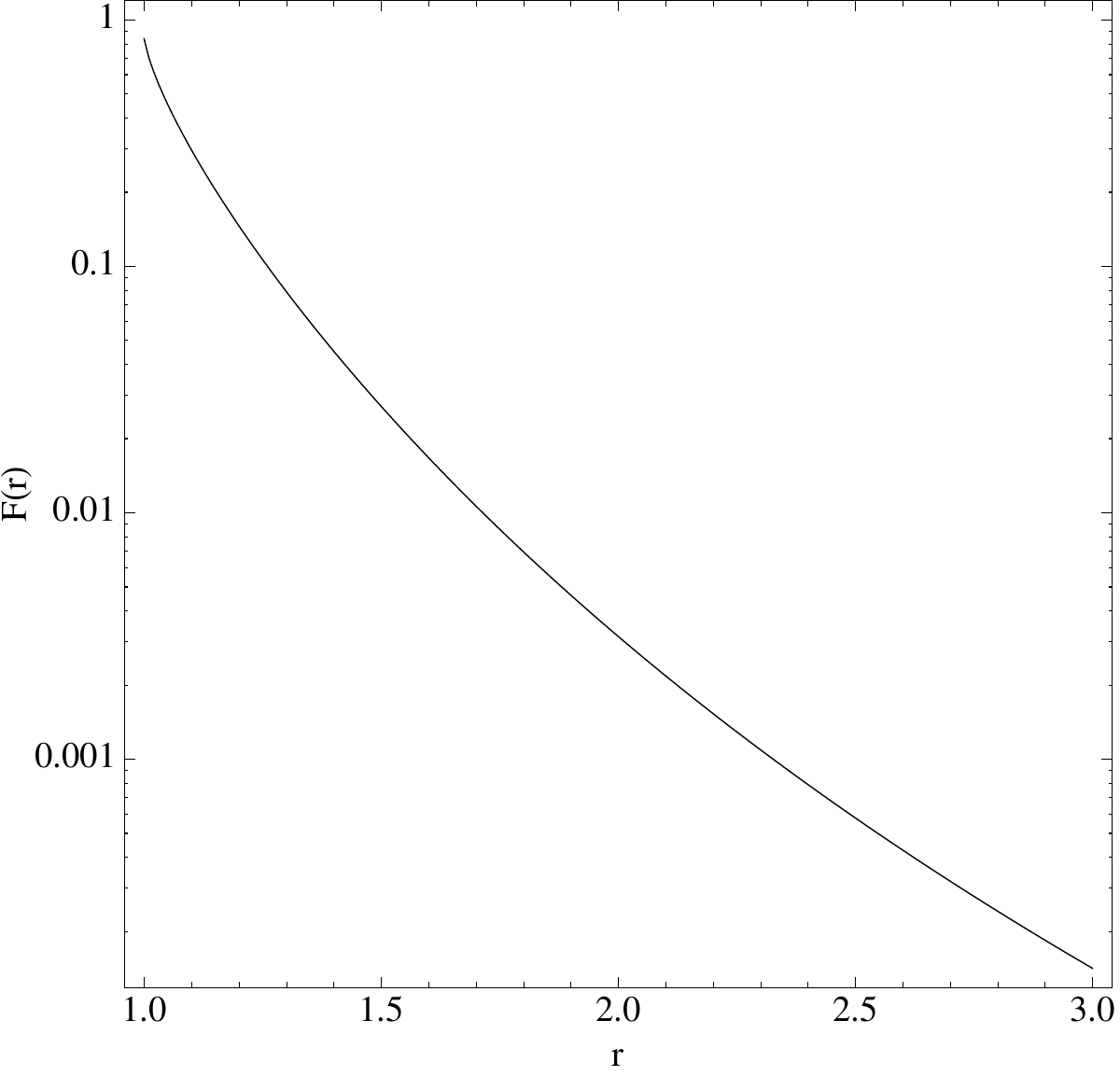}
\end{center}
\caption{Function of $F(r)$ of Eq.~(\ref{eq:total3body}) with $r =
  M_{\LH}/M_S$ (scalar DM) or $r=M_{\SH}/M_\chi$ (Majorana DM).}
\label{fig:fr}
\end{figure}

Concerning the 3-body cross section $\sgll$, we emphasize that the $r$
dependence is universal rewriting $\svgll$ as
\begin{equation}
  \label{eq:total3body}
\svgll = y_i^4\, {\alpha \over 32 \pi^2} \, {K_i\over M_i^2}\, F(r)
\end{equation}
with $y_i=g_l,y_l$ and $K_i =1/8,1$ for $i =
\chi,S$. Equation~(\ref{eq:total3body}) was obtained integrating the
amplitude (\ref{eq:amp3bodyS}) over the domain~\cite{Chen:1998dp}
$$
0 \leq x = {E_\gamma\over M_i} \leq 1 \quad \mbox{\rm and} \quad (1-x) \leq y = {E_f\over M_i} \leq 1.
$$ 
The function $F(r) $ is identical for the Majorana and scalar
cases and it is shown in Fig.~\ref{fig:fr}.
It behaves like $r^{-8}$, as expected, which is also the behaviour of
the 2-body cross section in the scalar case. On the contrary, in the
Majorana case the 2-body cross section behaves like $r^{-4}$. Notice
that a useful approximate expression of $\svgll$ has also been derived
previously in~\cite{Bringmann:2007nk} in the case of Majorana
dark matter. The latter reads
\begin{eqnarray}
\svgll&\approx& \frac{\alpha K_i y_i^4}{8 \pi^2 M_\chi^2}\left((r^2 + 1) \left(\pi^2/6 - \log\left[\frac{r^2 + 1}{2r^2}\right]^2 - 
       2 Li_2\left[\frac{r^2 + 1}{2 r^2}\right]\right)\right.\cr
 &&\hspace{2cm}\left.+ \frac{4 r^2 + 3}{r^2 + 
       1} + \frac{4 r^4 - 3 r^2 - 1}{2 r^2} \log\left[\frac{r^2 - 1}{r^2 + 1}\right]\right)\, ,
\end{eqnarray}
and agrees with our findings in the case of scalar dark matter using
$y_i=g_l,y_l$ and $K_i =1/8,1$ for $i = \chi,S$ as in our
Eq.~(\ref{eq:total3body}).

For the annihilation into $\gamma \gamma$, we follow the compact
formulation of~\cite{Tulin:2012uq} inspired by~\cite{Bergstrom:1989jr}
in the Majorana case and~\cite{Bertone:2009cb} in the scalar case. In
the notations used here, in the chiral limit, we thus made use of
\begin{eqnarray}
  \svgg&=&\frac{\alpha^2g_l^4}{256\pi^3M^2_{\chi}}{\cal I}(r)^2\cr
  &\mbox{with}&{\cal I}(r)=\int_0^1\frac{dx}{x}\log\left(\left|\frac{-x^2+(1-r^2)x+r^2}{x^2+(-1-r^2)x+r^2}\right|\right)
\end{eqnarray}
 for Majorana dark matter and 
 \begin{eqnarray}
   \svgg&=&\frac{\alpha^2 y_l^4}{64 \pi^3 M^2_S}|\mathcal{A}(r)|^2\cr
   &\mbox{with}&\mathcal{A}(r)=2-2\log\left[1-\frac{1}{r^2}\right]-2 r^2 \arcsin^2\left[\frac{1}{r}\right]
 \end{eqnarray}
for scalar dark matter. Notice though that $\svgg$ diverges in the
limit $r\rightarrow 1$ in the scalar case while it is under control in
the Majorana case~\cite{Bergstrom:1989jr}. In~\cite{Tulin:2012uq}, it is mentioned that the
approximation for $\svgg$ for scalar dark matter is actually expected
to break down in such a limit when $m_f\rightarrow 0$ is
considered. We leave for future work a more detailed analysis of
$\svgg$ in this framework.


 Finally, let us mention that we have obtained the results presented
 in Figs.~\ref{fig:sv3} and \ref{fig:yuk}  performing random scans over
 the parameter space:
\begin{eqnarray}
  1<&r&<5\cr
  10^{-3}<&y_l, g_l&<\pi\cr
  45 \,\GeV<&\mdm&< 10^4 \,\GeV
\end{eqnarray}
 setting the coupling of the scalar dark matter to zero and imposing
 that the dark matter relic abundance is $0.09<\Omega_{\rm dm }h^2<0.13$.
 Imposing that
 $y_f<\pi$ allow us to obtain viable dark matter candidates for $r<3
 (4)$ in the scalar (Majorana) dark matter case.

  \bibliography{bib3bdy}{}

\begin{thebibliography}{62}
\expandafter\ifx\csname natexlab\endcsname\relax\def\natexlab#1{#1}\fi
\expandafter\ifx\csname bibnamefont\endcsname\relax
  \def\bibnamefont#1{#1}\fi
\expandafter\ifx\csname bibfnamefont\endcsname\relax
  \def\bibfnamefont#1{#1}\fi
\expandafter\ifx\csname citenamefont\endcsname\relax
  \def\citenamefont#1{#1}\fi
\expandafter\ifx\csname url\endcsname\relax
  \def\url#1{\texttt{#1}}\fi
\expandafter\ifx\csname urlprefix\endcsname\relax\def\urlprefix{URL }\fi
\providecommand{\bibinfo}[2]{#2}
\providecommand{\eprint}[2][]{\url{#2}}

\bibitem[{\citenamefont{Bergstrom and Snellman}(1988)}]{Bergstrom:1988fp}
\bibinfo{author}{\bibfnamefont{L.}~\bibnamefont{Bergstrom}} \bibnamefont{and}
  \bibinfo{author}{\bibfnamefont{H.}~\bibnamefont{Snellman}},
  \bibinfo{journal}{Phys.Rev.} \textbf{\bibinfo{volume}{D37}},
  \bibinfo{pages}{3737} (\bibinfo{year}{1988}).

\bibitem[{\citenamefont{Rudaz}(1989)}]{Rudaz:1989ij}
\bibinfo{author}{\bibfnamefont{S.}~\bibnamefont{Rudaz}},
  \bibinfo{journal}{Phys.Rev.} \textbf{\bibinfo{volume}{D39}},
  \bibinfo{pages}{3549} (\bibinfo{year}{1989}).

\bibitem[{\citenamefont{Urban et~al.}(1992)\citenamefont{Urban, Bouquet,
  Degrange, Fleury, Kaplan et~al.}}]{Urban:1992ej}
\bibinfo{author}{\bibfnamefont{M.}~\bibnamefont{Urban}},
  \bibinfo{author}{\bibfnamefont{A.}~\bibnamefont{Bouquet}},
  \bibinfo{author}{\bibfnamefont{B.}~\bibnamefont{Degrange}},
  \bibinfo{author}{\bibfnamefont{P.}~\bibnamefont{Fleury}},
  \bibinfo{author}{\bibfnamefont{J.}~\bibnamefont{Kaplan}},
  \bibnamefont{et~al.}, \bibinfo{journal}{Phys.Lett.}
  \textbf{\bibinfo{volume}{B293}}, \bibinfo{pages}{149} (\bibinfo{year}{1992}),
  \eprint{hep-ph/9208255}.

\bibitem[{\citenamefont{Bringmann and Weniger}(2012)}]{Bringmann:2012ez}
\bibinfo{author}{\bibfnamefont{T.}~\bibnamefont{Bringmann}} \bibnamefont{and}
  \bibinfo{author}{\bibfnamefont{C.}~\bibnamefont{Weniger}},
  \bibinfo{journal}{Phys.Dark Univ.} \textbf{\bibinfo{volume}{1}},
  \bibinfo{pages}{194} (\bibinfo{year}{2012}), \eprint{1208.5481}.

\bibitem[{\citenamefont{$\rm
  Fermi-LAT~Collaboration$}(2013)}]{Fermi-LAT:2013uma}
\bibinfo{author}{\bibnamefont{$\rm Fermi-LAT~Collaboration$}}
  (\bibinfo{year}{2013}), \eprint{1305.5597}.

\bibitem[{\citenamefont{Abramowski et~al.}(2013)}]{Abramowski:2013ax}
\bibinfo{author}{\bibfnamefont{A.}~\bibnamefont{Abramowski}}
  \bibnamefont{et~al.} (\bibinfo{collaboration}{H.E.S.S. Collaboration}),
  \bibinfo{journal}{Phys.Rev.Lett.} \textbf{\bibinfo{volume}{110}},
  \bibinfo{pages}{041301} (\bibinfo{year}{2013}), \eprint{1301.1173}.

\bibitem[{\citenamefont{Bringmann et~al.}(2012)\citenamefont{Bringmann, Huang,
  Ibarra, Vogl, and Weniger}}]{Bringmann:2012vr}
\bibinfo{author}{\bibfnamefont{T.}~\bibnamefont{Bringmann}},
  \bibinfo{author}{\bibfnamefont{X.}~\bibnamefont{Huang}},
  \bibinfo{author}{\bibfnamefont{A.}~\bibnamefont{Ibarra}},
  \bibinfo{author}{\bibfnamefont{S.}~\bibnamefont{Vogl}}, \bibnamefont{and}
  \bibinfo{author}{\bibfnamefont{C.}~\bibnamefont{Weniger}},
  \bibinfo{journal}{JCAP} \textbf{\bibinfo{volume}{1207}}, \bibinfo{pages}{054}
  (\bibinfo{year}{2012}), \eprint{1203.1312}.

\bibitem[{\citenamefont{Weniger}(2012)}]{Weniger:2012tx}
\bibinfo{author}{\bibfnamefont{C.}~\bibnamefont{Weniger}},
  \bibinfo{journal}{JCAP} \textbf{\bibinfo{volume}{1208}}, \bibinfo{pages}{007}
  (\bibinfo{year}{2012}), \eprint{1204.2797}.

\bibitem[{\citenamefont{Dudas et~al.}(2012)\citenamefont{Dudas, Mambrini,
  Pokorski, and Romagnoni}}]{Dudas:2012pb}
\bibinfo{author}{\bibfnamefont{E.}~\bibnamefont{Dudas}},
  \bibinfo{author}{\bibfnamefont{Y.}~\bibnamefont{Mambrini}},
  \bibinfo{author}{\bibfnamefont{S.}~\bibnamefont{Pokorski}}, \bibnamefont{and}
  \bibinfo{author}{\bibfnamefont{A.}~\bibnamefont{Romagnoni}},
  \bibinfo{journal}{JHEP} \textbf{\bibinfo{volume}{1210}}, \bibinfo{pages}{123}
  (\bibinfo{year}{2012}), \eprint{1205.1520}.

\bibitem[{\citenamefont{Cline}(2012)}]{Cline:2012nw}
\bibinfo{author}{\bibfnamefont{J.~M.} \bibnamefont{Cline}},
  \bibinfo{journal}{Phys.Rev.} \textbf{\bibinfo{volume}{D86}},
  \bibinfo{pages}{015016} (\bibinfo{year}{2012}), \eprint{1205.2688}.

\bibitem[{\citenamefont{Ibarra et~al.}(2012)\citenamefont{Ibarra, Lopez~Gehler,
  and Pato}}]{Ibarra:2012dw}
\bibinfo{author}{\bibfnamefont{A.}~\bibnamefont{Ibarra}},
  \bibinfo{author}{\bibfnamefont{S.}~\bibnamefont{Lopez~Gehler}},
  \bibnamefont{and} \bibinfo{author}{\bibfnamefont{M.}~\bibnamefont{Pato}},
  \bibinfo{journal}{JCAP} \textbf{\bibinfo{volume}{1207}}, \bibinfo{pages}{043}
  (\bibinfo{year}{2012}), \eprint{1205.0007}.

\bibitem[{\citenamefont{Kyae and Park}(2013)}]{Kyae:2012vi}
\bibinfo{author}{\bibfnamefont{B.}~\bibnamefont{Kyae}} \bibnamefont{and}
  \bibinfo{author}{\bibfnamefont{J.-C.} \bibnamefont{Park}},
  \bibinfo{journal}{Phys.Lett.} \textbf{\bibinfo{volume}{B718}},
  \bibinfo{pages}{1425} (\bibinfo{year}{2013}), \eprint{1205.4151}.

\bibitem[{\citenamefont{Buckley and Hooper}(2012)}]{Buckley:2012ws}
\bibinfo{author}{\bibfnamefont{M.~R.} \bibnamefont{Buckley}} \bibnamefont{and}
  \bibinfo{author}{\bibfnamefont{D.}~\bibnamefont{Hooper}},
  \bibinfo{journal}{Phys.Rev.} \textbf{\bibinfo{volume}{D86}},
  \bibinfo{pages}{043524} (\bibinfo{year}{2012}), \eprint{1205.6811}.

\bibitem[{\citenamefont{Chu et~al.}(2012)\citenamefont{Chu, Hambye, Scarna, and
  Tytgat}}]{Chu:2012qy}
\bibinfo{author}{\bibfnamefont{X.}~\bibnamefont{Chu}},
  \bibinfo{author}{\bibfnamefont{T.}~\bibnamefont{Hambye}},
  \bibinfo{author}{\bibfnamefont{T.}~\bibnamefont{Scarna}}, \bibnamefont{and}
  \bibinfo{author}{\bibfnamefont{M.~H.} \bibnamefont{Tytgat}},
  \bibinfo{journal}{Phys.Rev.} \textbf{\bibinfo{volume}{D86}},
  \bibinfo{pages}{083521} (\bibinfo{year}{2012}), \eprint{1206.2279}.

\bibitem[{\citenamefont{Das et~al.}(2012)\citenamefont{Das, Ellwanger, and
  Mitropoulos}}]{Das:2012ys}
\bibinfo{author}{\bibfnamefont{D.}~\bibnamefont{Das}},
  \bibinfo{author}{\bibfnamefont{U.}~\bibnamefont{Ellwanger}},
  \bibnamefont{and}
  \bibinfo{author}{\bibfnamefont{P.}~\bibnamefont{Mitropoulos}},
  \bibinfo{journal}{JCAP} \textbf{\bibinfo{volume}{1208}}, \bibinfo{pages}{003}
  (\bibinfo{year}{2012}), \eprint{1206.2639}.

\bibitem[{\citenamefont{Weiner and Yavin}(2012)}]{Weiner:2012cb}
\bibinfo{author}{\bibfnamefont{N.}~\bibnamefont{Weiner}} \bibnamefont{and}
  \bibinfo{author}{\bibfnamefont{I.}~\bibnamefont{Yavin}},
  \bibinfo{journal}{Phys.Rev.} \textbf{\bibinfo{volume}{D86}},
  \bibinfo{pages}{075021} (\bibinfo{year}{2012}), \eprint{1206.2910}.

\bibitem[{\citenamefont{Tulin et~al.}(2013)\citenamefont{Tulin, Yu, and
  Zurek}}]{Tulin:2012uq}
\bibinfo{author}{\bibfnamefont{S.}~\bibnamefont{Tulin}},
  \bibinfo{author}{\bibfnamefont{H.-B.} \bibnamefont{Yu}}, \bibnamefont{and}
  \bibinfo{author}{\bibfnamefont{K.~M.} \bibnamefont{Zurek}},
  \bibinfo{journal}{Phys.Rev.} \textbf{\bibinfo{volume}{D87}},
  \bibinfo{pages}{036011} (\bibinfo{year}{2013}), \eprint{1208.0009}.

\bibitem[{\citenamefont{Acharya et~al.}(2012)\citenamefont{Acharya, Kane,
  Kumar, Lu, and Zheng}}]{Acharya:2012dz}
\bibinfo{author}{\bibfnamefont{B.~S.} \bibnamefont{Acharya}},
  \bibinfo{author}{\bibfnamefont{G.}~\bibnamefont{Kane}},
  \bibinfo{author}{\bibfnamefont{P.}~\bibnamefont{Kumar}},
  \bibinfo{author}{\bibfnamefont{R.}~\bibnamefont{Lu}}, \bibnamefont{and}
  \bibinfo{author}{\bibfnamefont{B.}~\bibnamefont{Zheng}}
  (\bibinfo{year}{2012}), \eprint{1205.5789}.

\bibitem[{\citenamefont{Gustafsson et~al.}(2007)\citenamefont{Gustafsson,
  Lundstrom, Bergstrom, and Edsjo}}]{Gustafsson:2007pc}
\bibinfo{author}{\bibfnamefont{M.}~\bibnamefont{Gustafsson}},
  \bibinfo{author}{\bibfnamefont{E.}~\bibnamefont{Lundstrom}},
  \bibinfo{author}{\bibfnamefont{L.}~\bibnamefont{Bergstrom}},
  \bibnamefont{and} \bibinfo{author}{\bibfnamefont{J.}~\bibnamefont{Edsjo}},
  \bibinfo{journal}{Phys.Rev.Lett.} \textbf{\bibinfo{volume}{99}},
  \bibinfo{pages}{041301} (\bibinfo{year}{2007}), \eprint{astro-ph/0703512}.

\bibitem[{\citenamefont{Mambrini}(2009)}]{Mambrini:2009ad}
\bibinfo{author}{\bibfnamefont{Y.}~\bibnamefont{Mambrini}},
  \bibinfo{journal}{JCAP} \textbf{\bibinfo{volume}{0912}}, \bibinfo{pages}{005}
  (\bibinfo{year}{2009}), \eprint{0907.2918}.

\bibitem[{\citenamefont{Jackson et~al.}(2010)\citenamefont{Jackson, Servant,
  Shaughnessy, Tait, and Taoso}}]{Jackson:2009kg}
\bibinfo{author}{\bibfnamefont{C.}~\bibnamefont{Jackson}},
  \bibinfo{author}{\bibfnamefont{G.}~\bibnamefont{Servant}},
  \bibinfo{author}{\bibfnamefont{G.}~\bibnamefont{Shaughnessy}},
  \bibinfo{author}{\bibfnamefont{T.~M.} \bibnamefont{Tait}}, \bibnamefont{and}
  \bibinfo{author}{\bibfnamefont{M.}~\bibnamefont{Taoso}},
  \bibinfo{journal}{JCAP} \textbf{\bibinfo{volume}{1004}}, \bibinfo{pages}{004}
  (\bibinfo{year}{2010}), \eprint{0912.0004}.

\bibitem[{\citenamefont{Arina et~al.}(2010)\citenamefont{Arina, Hambye, Ibarra,
  and Weniger}}]{Arina:2009uq}
\bibinfo{author}{\bibfnamefont{C.}~\bibnamefont{Arina}},
  \bibinfo{author}{\bibfnamefont{T.}~\bibnamefont{Hambye}},
  \bibinfo{author}{\bibfnamefont{A.}~\bibnamefont{Ibarra}}, \bibnamefont{and}
  \bibinfo{author}{\bibfnamefont{C.}~\bibnamefont{Weniger}},
  \bibinfo{journal}{JCAP} \textbf{\bibinfo{volume}{1003}}, \bibinfo{pages}{024}
  (\bibinfo{year}{2010}), \eprint{0912.4496}.

\bibitem[{\citenamefont{Schmidt-Hoberg
  et~al.}(2013)\citenamefont{Schmidt-Hoberg, Staub, and
  Winkler}}]{SchmidtHoberg:2012ip}
\bibinfo{author}{\bibfnamefont{K.}~\bibnamefont{Schmidt-Hoberg}},
  \bibinfo{author}{\bibfnamefont{F.}~\bibnamefont{Staub}}, \bibnamefont{and}
  \bibinfo{author}{\bibfnamefont{M.~W.} \bibnamefont{Winkler}},
  \bibinfo{journal}{JHEP} \textbf{\bibinfo{volume}{1301}}, \bibinfo{pages}{124}
  (\bibinfo{year}{2013}), \eprint{1211.2835}.

\bibitem[{\citenamefont{Wang and Han}(2013)}]{Wang:2012ts}
\bibinfo{author}{\bibfnamefont{L.}~\bibnamefont{Wang}} \bibnamefont{and}
  \bibinfo{author}{\bibfnamefont{X.-F.} \bibnamefont{Han}},
  \bibinfo{journal}{Phys.Rev.} \textbf{\bibinfo{volume}{D87}},
  \bibinfo{pages}{015015} (\bibinfo{year}{2013}), \eprint{1209.0376}.

\bibitem[{\citenamefont{Bergstrom}(1989)}]{Bergstrom:1989jr}
\bibinfo{author}{\bibfnamefont{L.}~\bibnamefont{Bergstrom}},
  \bibinfo{journal}{Phys.Lett.} \textbf{\bibinfo{volume}{B225}},
  \bibinfo{pages}{372} (\bibinfo{year}{1989}).

\bibitem[{\citenamefont{Flores et~al.}(1989)\citenamefont{Flores, Olive, and
  Rudaz}}]{Flores:1989ru}
\bibinfo{author}{\bibfnamefont{R.}~\bibnamefont{Flores}},
  \bibinfo{author}{\bibfnamefont{K.~A.} \bibnamefont{Olive}}, \bibnamefont{and}
  \bibinfo{author}{\bibfnamefont{S.}~\bibnamefont{Rudaz}},
  \bibinfo{journal}{Phys.Lett.} \textbf{\bibinfo{volume}{B232}},
  \bibinfo{pages}{377} (\bibinfo{year}{1989}).

\bibitem[{\citenamefont{Bringmann et~al.}(2008)\citenamefont{Bringmann,
  Bergstrom, and Edsjo}}]{Bringmann:2007nk}
\bibinfo{author}{\bibfnamefont{T.}~\bibnamefont{Bringmann}},
  \bibinfo{author}{\bibfnamefont{L.}~\bibnamefont{Bergstrom}},
  \bibnamefont{and} \bibinfo{author}{\bibfnamefont{J.}~\bibnamefont{Edsjo}},
  \bibinfo{journal}{JHEP} \textbf{\bibinfo{volume}{0801}}, \bibinfo{pages}{049}
  (\bibinfo{year}{2008}), \eprint{0710.3169}.

\bibitem[{\citenamefont{Goldberg}(1983)}]{Goldberg:1983nd}
\bibinfo{author}{\bibfnamefont{H.}~\bibnamefont{Goldberg}},
  \bibinfo{journal}{Phys.Rev.Lett.} \textbf{\bibinfo{volume}{50}},
  \bibinfo{pages}{1419} (\bibinfo{year}{1983}).

\bibitem[{\citenamefont{Baltz and Bergstrom}(2003)}]{Baltz:2002we}
\bibinfo{author}{\bibfnamefont{E.}~\bibnamefont{Baltz}} \bibnamefont{and}
  \bibinfo{author}{\bibfnamefont{L.}~\bibnamefont{Bergstrom}},
  \bibinfo{journal}{Phys.Rev.} \textbf{\bibinfo{volume}{D67}},
  \bibinfo{pages}{043516} (\bibinfo{year}{2003}), \eprint{hep-ph/0211325}.

\bibitem[{\citenamefont{Ciafaloni et~al.}(2011)\citenamefont{Ciafaloni,
  Cirelli, Comelli, De~Simone, Riotto et~al.}}]{Ciafaloni:2011sa}
\bibinfo{author}{\bibfnamefont{P.}~\bibnamefont{Ciafaloni}},
  \bibinfo{author}{\bibfnamefont{M.}~\bibnamefont{Cirelli}},
  \bibinfo{author}{\bibfnamefont{D.}~\bibnamefont{Comelli}},
  \bibinfo{author}{\bibfnamefont{A.}~\bibnamefont{De~Simone}},
  \bibinfo{author}{\bibfnamefont{A.}~\bibnamefont{Riotto}},
  \bibnamefont{et~al.}, \bibinfo{journal}{JCAP}
  \textbf{\bibinfo{volume}{1106}}, \bibinfo{pages}{018} (\bibinfo{year}{2011}),
  \eprint{1104.2996}.

\bibitem[{\citenamefont{Bell et~al.}(2011)\citenamefont{Bell, Dent, Galea,
  Jacques, Krauss et~al.}}]{Bell:2011if}
\bibinfo{author}{\bibfnamefont{N.~F.} \bibnamefont{Bell}},
  \bibinfo{author}{\bibfnamefont{J.~B.} \bibnamefont{Dent}},
  \bibinfo{author}{\bibfnamefont{A.~J.} \bibnamefont{Galea}},
  \bibinfo{author}{\bibfnamefont{T.~D.} \bibnamefont{Jacques}},
  \bibinfo{author}{\bibfnamefont{L.~M.} \bibnamefont{Krauss}},
  \bibnamefont{et~al.}, \bibinfo{journal}{Phys.Lett.}
  \textbf{\bibinfo{volume}{B706}}, \bibinfo{pages}{6} (\bibinfo{year}{2011}),
  \eprint{1104.3823}.

\bibitem[{\citenamefont{Barger et~al.}(2012)\citenamefont{Barger, Keung, and
  Marfatia}}]{Barger:2011jg}
\bibinfo{author}{\bibfnamefont{V.}~\bibnamefont{Barger}},
  \bibinfo{author}{\bibfnamefont{W.-Y.} \bibnamefont{Keung}}, \bibnamefont{and}
  \bibinfo{author}{\bibfnamefont{D.}~\bibnamefont{Marfatia}},
  \bibinfo{journal}{Phys.Lett.} \textbf{\bibinfo{volume}{B707}},
  \bibinfo{pages}{385} (\bibinfo{year}{2012}), \eprint{1111.4523}.

\bibitem[{\citenamefont{Garny et~al.}(2012)\citenamefont{Garny, Ibarra, and
  Vogl}}]{Garny:2011ii}
\bibinfo{author}{\bibfnamefont{M.}~\bibnamefont{Garny}},
  \bibinfo{author}{\bibfnamefont{A.}~\bibnamefont{Ibarra}}, \bibnamefont{and}
  \bibinfo{author}{\bibfnamefont{S.}~\bibnamefont{Vogl}},
  \bibinfo{journal}{JCAP} \textbf{\bibinfo{volume}{1204}}, \bibinfo{pages}{033}
  (\bibinfo{year}{2012}), \eprint{1112.5155}.

\bibitem[{\citenamefont{Bergstrom}(2012)}]{Bergstrom:2012bd}
\bibinfo{author}{\bibfnamefont{L.}~\bibnamefont{Bergstrom}},
  \bibinfo{journal}{Phys.Rev.} \textbf{\bibinfo{volume}{D86}},
  \bibinfo{pages}{103514} (\bibinfo{year}{2012}), \eprint{1208.6082}.

\bibitem[{\citenamefont{Boehm et~al.}(2004)\citenamefont{Boehm, Fayet, and
  Silk}}]{Boehm:2003ha}
\bibinfo{author}{\bibfnamefont{C.}~\bibnamefont{Boehm}},
  \bibinfo{author}{\bibfnamefont{P.}~\bibnamefont{Fayet}}, \bibnamefont{and}
  \bibinfo{author}{\bibfnamefont{J.}~\bibnamefont{Silk}},
  \bibinfo{journal}{Phys.Rev.} \textbf{\bibinfo{volume}{D69}},
  \bibinfo{pages}{101302} (\bibinfo{year}{2004}), \eprint{hep-ph/0311143}.

\bibitem[{\citenamefont{Fileviez~Perez and Wise}(2013)}]{Perez:2013nra}
\bibinfo{author}{\bibfnamefont{P.}~\bibnamefont{Fileviez~Perez}}
  \bibnamefont{and} \bibinfo{author}{\bibfnamefont{M.~B.} \bibnamefont{Wise}},
  \bibinfo{journal}{JHEP} \textbf{\bibinfo{volume}{1305}}, \bibinfo{pages}{094}
  (\bibinfo{year}{2013}), \eprint{1303.1452}.

\bibitem[{\citenamefont{Silveira and Zee}(1985)}]{Silveira:1985rk}
\bibinfo{author}{\bibfnamefont{V.}~\bibnamefont{Silveira}} \bibnamefont{and}
  \bibinfo{author}{\bibfnamefont{A.}~\bibnamefont{Zee}},
  \bibinfo{journal}{Phys.Lett.} \textbf{\bibinfo{volume}{B161}},
  \bibinfo{pages}{136} (\bibinfo{year}{1985}).

\bibitem[{\citenamefont{Veltman and Yndurain}(1989)}]{Veltman:1989vw}
\bibinfo{author}{\bibfnamefont{M.}~\bibnamefont{Veltman}} \bibnamefont{and}
  \bibinfo{author}{\bibfnamefont{F.}~\bibnamefont{Yndurain}},
  \bibinfo{journal}{Nucl.Phys.} \textbf{\bibinfo{volume}{B325}},
  \bibinfo{pages}{1} (\bibinfo{year}{1989}).

\bibitem[{\citenamefont{McDonald}(1994)}]{McDonald:1993ex}
\bibinfo{author}{\bibfnamefont{J.}~\bibnamefont{McDonald}},
  \bibinfo{journal}{Phys.Rev.} \textbf{\bibinfo{volume}{D50}},
  \bibinfo{pages}{3637} (\bibinfo{year}{1994}), \eprint{hep-ph/0702143}.

\bibitem[{\citenamefont{Burgess et~al.}(2001)\citenamefont{Burgess, Pospelov,
  and ter Veldhuis}}]{Burgess:2000yq}
\bibinfo{author}{\bibfnamefont{C.}~\bibnamefont{Burgess}},
  \bibinfo{author}{\bibfnamefont{M.}~\bibnamefont{Pospelov}}, \bibnamefont{and}
  \bibinfo{author}{\bibfnamefont{T.}~\bibnamefont{ter Veldhuis}},
  \bibinfo{journal}{Nucl.Phys.} \textbf{\bibinfo{volume}{B619}},
  \bibinfo{pages}{709} (\bibinfo{year}{2001}), \eprint{hep-ph/0011335}.

\bibitem[{\citenamefont{Boehm and Uwer}(2006)}]{Boehm:2006df}
\bibinfo{author}{\bibfnamefont{C.}~\bibnamefont{Boehm}} \bibnamefont{and}
  \bibinfo{author}{\bibfnamefont{P.}~\bibnamefont{Uwer}}
  (\bibinfo{year}{2006}), \eprint{hep-ph/0606058}.

\bibitem[{\citenamefont{Peskin and Schroeder}(1995)}]{Peskin:1995ev}
\bibinfo{author}{\bibfnamefont{M.~E.} \bibnamefont{Peskin}} \bibnamefont{and}
  \bibinfo{author}{\bibfnamefont{D.~V.} \bibnamefont{Schroeder}}
  (\bibinfo{year}{1995}).

\bibitem[{\citenamefont{Djouadi}(2008)}]{Djouadi:2005gi}
\bibinfo{author}{\bibfnamefont{A.}~\bibnamefont{Djouadi}},
  \bibinfo{journal}{Phys.Rept.} \textbf{\bibinfo{volume}{457}},
  \bibinfo{pages}{1} (\bibinfo{year}{2008}), \eprint{hep-ph/0503172}.

\bibitem[{\citenamefont{Bertone et~al.}(2009)\citenamefont{Bertone, Jackson,
  Shaughnessy, Tait, and Vallinotto}}]{Bertone:2009cb}
\bibinfo{author}{\bibfnamefont{G.}~\bibnamefont{Bertone}},
  \bibinfo{author}{\bibfnamefont{C.}~\bibnamefont{Jackson}},
  \bibinfo{author}{\bibfnamefont{G.}~\bibnamefont{Shaughnessy}},
  \bibinfo{author}{\bibfnamefont{T.~M.} \bibnamefont{Tait}}, \bibnamefont{and}
  \bibinfo{author}{\bibfnamefont{A.}~\bibnamefont{Vallinotto}},
  \bibinfo{journal}{Phys.Rev.} \textbf{\bibinfo{volume}{D80}},
  \bibinfo{pages}{023512} (\bibinfo{year}{2009}), \eprint{0904.1442}.

\bibitem[{\citenamefont{Garny et~al.}(2013)\citenamefont{Garny, Ibarra, Pato,
  and Vogl}}]{Garny:2013ama}
\bibinfo{author}{\bibfnamefont{M.}~\bibnamefont{Garny}},
  \bibinfo{author}{\bibfnamefont{A.}~\bibnamefont{Ibarra}},
  \bibinfo{author}{\bibfnamefont{M.}~\bibnamefont{Pato}}, \bibnamefont{and}
  \bibinfo{author}{\bibfnamefont{S.}~\bibnamefont{Vogl}}
  (\bibinfo{year}{2013}), \eprint{1306.6342}.

\bibitem[{\citenamefont{Srednicki et~al.}(1988)\citenamefont{Srednicki,
  Watkins, and Olive}}]{Srednicki:1988ce}
\bibinfo{author}{\bibfnamefont{M.}~\bibnamefont{Srednicki}},
  \bibinfo{author}{\bibfnamefont{R.}~\bibnamefont{Watkins}}, \bibnamefont{and}
  \bibinfo{author}{\bibfnamefont{K.~A.} \bibnamefont{Olive}},
  \bibinfo{journal}{Nucl.Phys.} \textbf{\bibinfo{volume}{B310}},
  \bibinfo{pages}{693} (\bibinfo{year}{1988}).

\bibitem[{\citenamefont{Gondolo and Gelmini}(1991)}]{Gondolo:1990dk}
\bibinfo{author}{\bibfnamefont{P.}~\bibnamefont{Gondolo}} \bibnamefont{and}
  \bibinfo{author}{\bibfnamefont{G.}~\bibnamefont{Gelmini}},
  \bibinfo{journal}{Nucl.Phys.} \textbf{\bibinfo{volume}{B360}},
  \bibinfo{pages}{145} (\bibinfo{year}{1991}).

\bibitem[{\citenamefont{Kolb and Turner}(1990)}]{Kolb:1990vq}
\bibinfo{author}{\bibfnamefont{E.~W.} \bibnamefont{Kolb}} \bibnamefont{and}
  \bibinfo{author}{\bibfnamefont{M.~S.} \bibnamefont{Turner}},
  \bibinfo{journal}{Front.Phys.} \textbf{\bibinfo{volume}{69}},
  \bibinfo{pages}{1} (\bibinfo{year}{1990}).

\bibitem[{\citenamefont{Griest and Seckel}(1991)}]{Griest:1990kh}
\bibinfo{author}{\bibfnamefont{K.}~\bibnamefont{Griest}} \bibnamefont{and}
  \bibinfo{author}{\bibfnamefont{D.}~\bibnamefont{Seckel}},
  \bibinfo{journal}{Phys.Rev.} \textbf{\bibinfo{volume}{D43}},
  \bibinfo{pages}{3191} (\bibinfo{year}{1991}).

\bibitem[{\citenamefont{Belanger et~al.}(2011)\citenamefont{Belanger, Boudjema,
  Brun, Pukhov, Rosier-Lees et~al.}}]{Belanger:2010gh}
\bibinfo{author}{\bibfnamefont{G.}~\bibnamefont{Belanger}},
  \bibinfo{author}{\bibfnamefont{F.}~\bibnamefont{Boudjema}},
  \bibinfo{author}{\bibfnamefont{P.}~\bibnamefont{Brun}},
  \bibinfo{author}{\bibfnamefont{A.}~\bibnamefont{Pukhov}},
  \bibinfo{author}{\bibfnamefont{S.}~\bibnamefont{Rosier-Lees}},
  \bibnamefont{et~al.}, \bibinfo{journal}{Comput.Phys.Commun.}
  \textbf{\bibinfo{volume}{182}}, \bibinfo{pages}{842} (\bibinfo{year}{2011}),
  \eprint{1004.1092}.

\bibitem[{\citenamefont{Christensen and Duhr}(2009)}]{Christensen:2008py}
\bibinfo{author}{\bibfnamefont{N.~D.} \bibnamefont{Christensen}}
  \bibnamefont{and} \bibinfo{author}{\bibfnamefont{C.}~\bibnamefont{Duhr}},
  \bibinfo{journal}{Comput.Phys.Commun.} \textbf{\bibinfo{volume}{180}},
  \bibinfo{pages}{1614} (\bibinfo{year}{2009}), \eprint{0806.4194}.

\bibitem[{\citenamefont{Drees and Nojiri}(1993)}]{Drees:1993bu}
\bibinfo{author}{\bibfnamefont{M.}~\bibnamefont{Drees}} \bibnamefont{and}
  \bibinfo{author}{\bibfnamefont{M.}~\bibnamefont{Nojiri}},
  \bibinfo{journal}{Phys.Rev.} \textbf{\bibinfo{volume}{D48}},
  \bibinfo{pages}{3483} (\bibinfo{year}{1993}), \eprint{hep-ph/9307208}.

\bibitem[{\citenamefont{Aprile et~al.}(2012)}]{Aprile:2012nq}
\bibinfo{author}{\bibfnamefont{E.}~\bibnamefont{Aprile}} \bibnamefont{et~al.}
  (\bibinfo{collaboration}{XENON100 Collaboration}),
  \bibinfo{journal}{Phys.Rev.Lett.} \textbf{\bibinfo{volume}{109}},
  \bibinfo{pages}{181301} (\bibinfo{year}{2012}), \eprint{1207.5988}.

\bibitem[{\citenamefont{Cline et~al.}(2013)\citenamefont{Cline, Kainulainen,
  Scott, and Weniger}}]{Cline:2013gha}
\bibinfo{author}{\bibfnamefont{J.~M.} \bibnamefont{Cline}},
  \bibinfo{author}{\bibfnamefont{K.}~\bibnamefont{Kainulainen}},
  \bibinfo{author}{\bibfnamefont{P.}~\bibnamefont{Scott}}, \bibnamefont{and}
  \bibinfo{author}{\bibfnamefont{C.}~\bibnamefont{Weniger}}
  (\bibinfo{year}{2013}), \eprint{1306.4710}.

\bibitem[{\citenamefont{Deshpande and Ma}(1978)}]{Deshpande:1977rw}
\bibinfo{author}{\bibfnamefont{N.~G.} \bibnamefont{Deshpande}}
  \bibnamefont{and} \bibinfo{author}{\bibfnamefont{E.}~\bibnamefont{Ma}},
  \bibinfo{journal}{Phys.Rev.} \textbf{\bibinfo{volume}{D18}},
  \bibinfo{pages}{2574} (\bibinfo{year}{1978}).

\bibitem[{\citenamefont{Ma}(2006)}]{Ma:2006km}
\bibinfo{author}{\bibfnamefont{E.}~\bibnamefont{Ma}},
  \bibinfo{journal}{Phys.Rev.} \textbf{\bibinfo{volume}{D73}},
  \bibinfo{pages}{077301} (\bibinfo{year}{2006}), \eprint{hep-ph/0601225}.

\bibitem[{\citenamefont{Barbieri et~al.}(2006)\citenamefont{Barbieri, Hall, and
  Rychkov}}]{Barbieri:2006dq}
\bibinfo{author}{\bibfnamefont{R.}~\bibnamefont{Barbieri}},
  \bibinfo{author}{\bibfnamefont{L.~J.} \bibnamefont{Hall}}, \bibnamefont{and}
  \bibinfo{author}{\bibfnamefont{V.~S.} \bibnamefont{Rychkov}},
  \bibinfo{journal}{Phys.Rev.} \textbf{\bibinfo{volume}{D74}},
  \bibinfo{pages}{015007} (\bibinfo{year}{2006}), \eprint{hep-ph/0603188}.

\bibitem[{\citenamefont{Garcia-Cely and Ibarra}(2013)}]{Garcia-Cely:2013zga}
\bibinfo{author}{\bibfnamefont{C.}~\bibnamefont{Garcia-Cely}} \bibnamefont{and}
  \bibinfo{author}{\bibfnamefont{A.}~\bibnamefont{Ibarra}}
  (\bibinfo{year}{2013}), \eprint{1306.4681}.

\bibitem[{\citenamefont{Cheung and Seto}(2004)}]{Cheung:2004xm}
\bibinfo{author}{\bibfnamefont{K.}~\bibnamefont{Cheung}} \bibnamefont{and}
  \bibinfo{author}{\bibfnamefont{O.}~\bibnamefont{Seto}},
  \bibinfo{journal}{Phys.Rev.} \textbf{\bibinfo{volume}{D69}},
  \bibinfo{pages}{113009} (\bibinfo{year}{2004}), \eprint{hep-ph/0403003}.

\bibitem[{\citenamefont{Ishiwata and Wise}(2013)}]{Ishiwata:2013gma}
\bibinfo{author}{\bibfnamefont{K.}~\bibnamefont{Ishiwata}} \bibnamefont{and}
  \bibinfo{author}{\bibfnamefont{M.~B.} \bibnamefont{Wise}}
  (\bibinfo{year}{2013}), \eprint{1307.1112}.

\bibitem[{\citenamefont{Toma}(2013)}]{Toma:2013bka}
\bibinfo{author}{\bibfnamefont{T.}~\bibnamefont{Toma}} (\bibinfo{year}{2013}),
  \eprint{1307.6181}.

\bibitem[{\citenamefont{Chen and Kamionkowski}(1998)}]{Chen:1998dp}
\bibinfo{author}{\bibfnamefont{X.-l.} \bibnamefont{Chen}} \bibnamefont{and}
  \bibinfo{author}{\bibfnamefont{M.}~\bibnamefont{Kamionkowski}},
  \bibinfo{journal}{JHEP} \textbf{\bibinfo{volume}{9807}}, \bibinfo{pages}{001}
  (\bibinfo{year}{1998}), \eprint{hep-ph/9805383}.

\end{thebibliography}

\end{document}